# Finding topological subgraphs is fixed-parameter tractable

Martin Grohe
Institut für Informatik
Humboldt-Universität zu Berlin
Unter den Linden 6,
10099 Berlin, Germany
grohe@informatik.hu-berlin.de

Ken-ichi Kawarabayashi [*]
National Institute of Informatics
2-1-2 Hitotsubashi, Chiyoda-ku
Tokyo 101-8430, Japan
k_keniti@nii.ac.jp

Dániel Marx
Institut für Informatik
Humboldt-Universität zu Berlin
Unter den Linden 6,
10099 Berlin, Germany
dmarx@informatik.hu-berlin.de

Paul Wollan
Department of Computer Science
University of Rome, *La Sapienza*
Via Salaria 113
Rome, 00198 Italy
wollan@di.uniroma1.it

**Abstract**

We show that for every fixed undirected graph $H$, there is a $O(|V(G)|^3)$ time algorithm that tests, given a graph $G$, if $G$ contains $H$ as a topological subgraph (that is, a subdivision of $H$ is subgraph of $G$). This shows that topological subgraph testing is fixed-parameter tractable, resolving a longstanding open question of Downey and Fellows from 1992.

As a corollary, for every $H$ we obtain an $O(|V(G)|^3)$ time algorithm that tests if there is an immersion of $H$ into a given graph $G$. This answers another open question raised by Downey and Fellows in 1992.

## 1 Introduction

A graph $H$ is a *topological subgraph* (or *topological minor*) of graph $G$ if a subdivision of $H$ is a subgraph of $G$. Equivalently, $H$ is a topological subgraph of $G$ if $H$ can be obtained from $G$ by deleting edges, deleting vertices, and dissolving degree 2 vertices (which means deleting the vertex and making its two neighbors adjacent). This notion appears for example in the classical result of Kuratowski in 1935 stating that a graph is planar if and only if it does not have a topological subgraph isomorphic to $K_5$ or $K_{3,3}$.

Given graphs $H$ and $G$, it is NP-complete to decide if $H$ is a topological subgraph of $G$ (e.g., a cycle of length $|V(G)|$ is a topological subgraph of $G$ if and only if $G$ is Hamiltonian). On the other hand, our main result shows that for every fixed $H$, there is a cubic algorithm:

**Theorem 1.1.** *For every fixed graph $H$, there is a $O(|V(G)|^3)$ time algorithm that decides if $H$ is a topological subgraph of $G$.*

Actually, our algorithm is uniform in $H$, and this shows that the problem of testing if $H$ is a topological subgraph of $G$ is fixed-parameter tractable parameterized by the number of vertices of $H$. Recall that a problem is *fixed-parameter tractable* by some parameter $k$ if it can be solved in time $f(k) \cdot n^{O(1)}$ for a function $f$ depending only on $k$. Thus Theorem 1.1 answers a longstanding open question, first raised in 1992 by Downey and Fellows [3] and then restated at many places, including the open problem list of the monograph [4]. The problem of testing for topological subgraphs, which is also known as the subgraph homeomorphism problem, was already studied in the 1970s by Lapaugh and Rivest [10] (also see [7]). Fortune, Hopcroft, and Wyllie [6] studied the directed version of the problem and showed that there are simple digraphs $H$ such that the problem of testing whether a given digraph $G$ contains $H$ as a (directed) topological subgraph is NP-complete. In a major breakthrough, Robertson and Seymour [11] proved that this cannot happen for undirected graphs: For every (undirected) graph $H$ there is a polynomial time algorithm testing whether a given graph $G$ contains $H$ as a topological subgraph. (We will discuss Robertson and Seymour's result in more detail below.) However, the running time of Robertson and Seymour's algorithm is $|V(G)|^{|V(H)|}$. This prompted Downey and Fellows' questions of whether the problem is fixed-parameter tractable. Our Theorem 1 answers this question.

---

[*]Research partly supported by Japan Society for the Promotion of Science, Grant-in-Aid for Scientific Research, by Kayamori Foundation and by Inoue Research Award for Young Scientists.

We also study the related problem of testing for immersed subgraphs. An *immersion* of a graph $H$ into a graph $G$ is defined like a topological embedding, expect that the paths in $G$ corresponding to the edges of $H$ are only required to be edge disjoint instead of internally vertex disjoint. Formally, an immersion of $H$ into $G$ is a mapping $\alpha$ that associates with each vertex $v \in V(H)$ a vertex $\alpha(v) \in V(G)$ and with each edge $e = vw \in E(H)$ a path $\alpha(e)$ in $G$ with endpoints $\alpha(v)$ and $\alpha(w)$ in such a way that the paths $\alpha(e)$ for $e \in E(H)$ are mutually edge disjoint. Robertson and Seymour [14] showed that graphs are well-quasi-ordered under the immersion relation, proving a conjecture of Nash-Williams. Here we obtain the following algorithmic result as a corollary to Theorem 1.1:

**Corollary 1.2.** *For every fixed graph $H$, there is a $O(|V(G)|^3)$ time algorithm that decides if there is an immersion of $H$ into $G$.*

Again, our algorithm is uniform in $H$, which implies that the immersion problem is fixed-parameter tractable. This answers another open question by Downey and Fellows [3, 4].

Yet another related problem is minor containment testing. We say that graph $H$ is a *minor* of $G$ if $H$ can be obtained from $G$ by deleting vertices, deleting edges, and contracting edges. A celebrated result of Robertson and Seymour [11] shows that for every fixed $H$, there is a $O(|V(G)|^3)$ time algorithm for testing if $H$ is a minor of $G$. Their algorithm actually solves a more general rooted version of the problem. This rooted version contains as a special case the $k$-DISJOINT PATHS problem, where given pairs $(s_1, t_1)$, ..., $(s_k, t_k)$ of vertices, the task is to find vertex disjoint paths $P_1$, ..., $P_k$ such that $P_i$ connects $s_i$ and $t_i$. It is not difficult to reduce testing if $H$ is a topological subgraph of $G$ to $k$-DISJOINT PATHS. For each vertex $v$ of $H$, we guess a vertex $v'$ of $G$, and then for each edge $uv$ of $H$, we find a path connecting $u'$ and $v'$ in $G$ such that these $|E(H)|$ paths are pairwise internally disjoint. This approach yields the $|V(G)|^{O(|V(H)|)}$ time algorithm for topological subgraph testing mentioned above.

Our algorithm for finding topological subgraphs follows the general framework of Robertson and Seymour for minor testing, but it deviates from it significantly. Let us give a very high-level overview of Robertson and Seymour's algorithm [11]. If the treewidth of $G$ is "small," then standard techniques allow us to solve the problem in linear time. If the treewidth of $G$ is "large," then we find an *irrelevant vertex* whose deletion provably does not change the answer to the problem. By iteratively finding and deleting irrelevant vertices, we eventually arrive to a $G$ whose treewidth is small. To find an irrelevant vertex if the treewidth of $G$ is large, we use the the so-called *Weak Structure Theorem*, which allows us either to find a large clique minor or to show that the graph has a large "flat wall.". The case of a large clique minor is easy to handle: if there are no roots, then it immediately solves the problem (as every small graph appears in the large clique minor) and even if roots are present, we can argue that a large part of the clique is irrelevant. The most difficult part of the algorithm is to deal with the case of a flat wall and to identify an irrelevant vertex there. Indeed, this case needs the majority of the work. The analysis of this case requires the whole series of Graph Minor papers and the structure theorem of [12]. Very recently, a significantly simpler treatment of this case was presented in [9].

Let us now give an overview for our algorithm. The case of small treewidth goes through for topological minor testing without any difficulty. The new proof in [9] for minor testing in the case when there is no large clique minor can be adapted for topological minor testing. Specifically, for the case where there is a large flat wall, using the unique linkage theorem [13] and its much shorter proof [9], we can indeed find an irrelevant vertex in the middle of the large flat wall. This case is similar to that for the minor testing, however, we may need to change almost all of the branch vertices of a given topological minor inside the flat wall. This gives rise to some amount of technical difficulties, which we overcome in this paper. Let us emphasis that our proof of the correctness for our algorithm does not depend on the full power of the graph minor structure theorem [12], while Robertson and Seymour's analysis for their algorithm do needs the whole series of Graph Minor papers and the structure theorem of [12]. Utilizing some results in [9], we are able to avoid the much of the heavy machinery of the graph minor structure theory.

Let us now look at the case when there is a large clique minor. Identifying a large clique minor was an easy situation to handle in the case of finding minors, but it is not obvious how it is of any use in the case of finding topological subgraphs. The problem is that the degrees of the vertices matter much more in finding topological subgraphs than in finding minors. If $H$ is, say, 4-regular and we have found a large clique minor in a part of $G$ that contains only degree-3 vertices, then this clique minor does not immediately solve the problem. Furthermore, as $G$ can contain many vertices of degree at least 4 close to this clique minor and each such vertex is potentially the image of some vertex of $H$, there is no easy argument that shows that some part of the clique is irrelevant. We circumvent these problems by introducing a new operation that was not present in the framework of [11]. If a small number of vertices can separate away a large part of the graph, then we recursively "understand" this part and then replace it with an equivalent smaller graph. We show that if no such step can be performed, then we can completely understand how the large clique minor can be used by a topological subgraph. This new operation and the associated recursion changes the high-level structure of our algorithm considerably: unlike in [11], it is no longer just an iterative removal of irrelevant vertices.

Similarly to [11], we define and solve a very general rooted version of the problem ("finding folios"). It is important to point out that we are solving this rooted generalization not (only) for the sake of obtaining maximum generality of the result. In the recursion steps involving separators, we argue about topological subgraphs using the separator in a certain way, and the concept of roots is needed to express these requirements.

## 2 Folios

A *rooted graph* is an undirected graph $G$ with a set $R(G) \subseteq V(G)$ of vertices specified as roots and an injective mapping $\rho_G : R(G) \to \mathbb{N}$ assigning a distinct positive integer label to each root vertex. Isomorphism of rooted graphs are defined the obvious way, i.e., roots must be mapped to roots with the same label. We say that two rooted graphs $G_1$ and $G_2$ are *compatible* if $\rho_{G_1}(R(G_1)) = \rho_{G_2}(R(G_2))$, i.e. the same set of positive integers appear on $G_1$ and $G_2$ (which means in particular that $|R(G_1)| = |R(G_2)|$).

We say that rooted graph $H$ is a *topological minor* of rooted graph $G$ if there is a mapping $\phi$ (a *model* of $H$ in $G$) that assigns to each $v \in V(H)$ a vertex $\phi(v) \in V(G)$ and to each $e \in E(G)$ a path $\phi(e)$ in $G$ such that

(1) The vertices $\phi(v)$ ($v \in V(H)$) are distinct.
(2) If $u, v \in V(H)$ are the endpoints of $e \in E(H)$, then path $\phi(e)$ connects $\phi(u)$ and $\phi(v)$.
(3) The paths $\phi(e)$ ($e \in E(H)$) are pairwise internally vertex disjoint, i.e., the internal vertices of $\phi(e)$ do not appear as an (internal or end) vertex of $\phi(e')$ for any $e' \neq e$.
(4) For every $v \in R(H)$, $\rho_G(\phi(v)) = \rho_H(v)$.

Even if $H$ is a topological minor of $G$, they are not necessarily compatible: $G$ can have more root vertices than $H$.

The *folio* of $G$ is the set of all topological minors of $G$. Clearly, the folio is closed under isomorphism, i.e., if rooted graphs $H$ and $H'$ are isomorphic and $H$ is in the folio of $G$, then $H'$ is in the folio as well. If $\delta \geq 0$ is an integer, then the $\delta$-*folio* of $G$ contains every topological minor $H$ of $G$ with $|E(H)| + \text{is}(H) \leq \delta$, where $\text{is}(H)$ is the number of isolated vertices of $H$. Obviously, every graph in the $\delta$-folio has at most $2\delta$ vertices.

**Observation 2.1.** *The number of distinct graphs (up to isomorphism) in the $\delta$-folio of $G$ can be bounded by a function of $\delta$ and $|R(G)|$.*

There are $2^{\binom{|R(G)|}{2}}$ possible undirected graphs on $R(G)$. For each such graph $X$, we slightly abuse notation by defining $G + X$ the obvious way. The rooted graph $G + X$ has a $\delta$-folio, which may or may not be different from the $\delta$-folio of $G$. The $2^{\binom{|R(G)|}{2}}$-tuple of all these $\delta$-folios will be called the *extended $\delta$-folio* of $G$.

Given an extended $\delta$-folio $\mathcal{F}$, a *representative* of $\mathcal{F}$ is a rooted graph $G$ whose extended $\delta$-folio is $\mathcal{F}$. We define the constant $L_{\delta,r}$ to be the smallest integer such that for every rooted graph $G$ with at most $r$ roots, the extended $\delta$-folio of $G$ has a representative on at most $L_{\delta,r}$ vertices. It is clear that $L_{\delta,r}$ is finite.

**Lemma 2.2.** *There is a computable function $\ell(\delta, r)$ with $L_{\delta,r} \leq \ell(\delta, r)$ for every $\delta, r \geq 0$.*

*Proof.* Let $G$ be a graph of minimum order with a given $\delta$-folio with $r$ roots. Suppose that the $\delta$-folio consists of graphs $H_1, \ldots, H_m$. For each $i \in [m]$, let $H_i'$ be a subgraph of $G$ isomorphic to a subdivision of $H_i$. Let $W_i \subseteq V(H_i')$ be the set of vertices corresponding to the vertices of $H_i$, and let $\mathcal{P}_i$ be the set of paths in $H_i'$ corresponding to the edges of $H_i$. Note that $G = \bigcup_{i=1}^m H_i'$ by the minimality of $G$. Let $W := \bigcup_{i=1}^m W_i$ and $\mathcal{P} := \bigcup_{i=1}^m \mathcal{P}_i$. In the terminology of Section 5, $\mathcal{P}$ is a path system. Let $k := |\mathcal{P}|$; obviously $k$ can be effectively bounded in terms of $\delta$ and $r$. It follows from the Unique Linkage Theorem 5.11, or more precisely Corollary 5.12, and the minimality of $G$ that the treewidth of $G$ is bounded by $w(k)$ for some computable function $w$. The computability of $w$ can be checked by going through the proof of the Unique Linkage Theorem [13, 9].

It is easy to see that the folio of a rooted graph can be defined in monadic-second order logic MSO. That is, we can find an MSO-sentence $\varphi$ stating that a graph has the desired folio. Hence we only have to prove that an MSO-sentence $\varphi$ that has a model of treewidth at most $\ell$ has such a model of order at most $g(\varphi, \ell)$, for some computable function $g$. This is well-known. To prove it, we give a translation (an *MSO-transduction*) that transforms $\varphi$ to a sentence $\varphi^*$ in the language of colored trees and associates with every graph $G$ of treewidth at most $\ell$ and every tree decomposition $D$ of $G$ of width at most $\ell$ a colored tree $T(G,D)$ of roughly the same size as $G$ such that $\varphi^*$ is only satisfied by trees of the form $T(G,D)$, and $T(G,D)$ satisfies $\varphi^*$ if and only if $G$ satisfies $\varphi$ (such a translation is described, for example, in Section 11.4 of [5]). Then we use a theorem due Thatcher and Wright [16] to (effectively) construct a tree automaton $A$ that accepts a tree $T$ if and only if $T$ satisfies $\varphi^*$. By a Pumping-Lemma argument, we obtain an effective bound $h(A)$ such that if $A$ accepts any tree at all, then it accepts a tree of size at most $h(A)$. As all the transformations involved are computable, this yields the desired bound on the smallest model of $\varphi$ and thus on the size of the smallest graph with the given folio. □

The (extended) $\delta$-folio of a graph $G$ *with respect to* a set $Z \subseteq V(G)$ is the (extended) $\delta$-folio of the graph $G'$, where $G'$ has the same set of vertices and edges as $G$, but $R(G') = Z$. We will use this notion to avoid defining new graphs that differ only in the set of roots. Some straightforward observations:

**Proposition 2.3.** *Let $G$ be a rooted graph and let $\delta \geq 0$ be an integer.*

*(1) The extended 0-folio of $G$ contains only the empty graph.*
*(2) Let $R \subseteq Q \subseteq V(G)$ be two sets of vertices. The $\delta$-folio of $G$ with respect to $R$ can be obtained from the $\delta$-folio of $G$ with respect to $Q$.*
*(3) Let $R_1$, …, $R_t$ be subsets of $V(G)$ such that for every subset $Q \subseteq R(G)$ of size at most $2\delta$ there is a $1 \leq i \leq t$ such that $Q \subseteq R_i$. The $\delta$-folio of $G$ can be obtained from the $\delta$-folios of $G$ with respect to $R_1$, …, $R_t$.*
*(4) The extended $\delta$-folio of $G$ can be obtained from the $(\delta + |R(G)|)$-folio of $G$.*

## 2.1 Separations and replacements

A *separation* of a graph $G$ is a pair $(A, B)$ of subgraphs such that $V(G) = V(A) \cup V(B)$, $E(G) = E(A) \cup E(B)$, and $E(A) \cap E(B) = \emptyset$. The *order* of the separation $(A, B)$ is $|V(A) \cap V(B)|$.

Let $(A, B)$ be a separation of rooted graph $G$ such that $V(A) \cap V(B) \subseteq R(G)$. Let $A'$ be a rooted graph compatible with $A$. *Replacing $A$ with $A'$ in the separation* $(A, B)$ gives the graph $G'$ defined as follows. We have $V(G') = V(A') \cup (V(B) \setminus V(A))$, $G'$ has every edge of $A'$ and $B \setminus V(A)$, and $G'$ has the following additional edges: if $u \in V(A) \cap V(B)$ and $v \in V(B) \setminus V(A)$ are adjacent in $G$, and $u' \in V(A')$ is a vertex with $\rho_A(u) = \rho_{A'}(u')$, then $u'$ and $v$ are adjacent in $G'$. Intuitively, we remove $A$ from $G$, and replace it by $A'$ such that the role of $V(A) \cap V(B)$ is taken by the matching root vertices of $A'$. The following lemmas show how the folio changes after replacement:

**Lemma 2.4.** *Let $(G_1, G_2)$ be a separation of a rooted graph $G$, let $S = V(G_1) \cap V(G_2)$, and suppose that $S \subseteq R(G)$. Let $G_1'$ be a rooted graph compatible with $G_1$ such that $G_1$ and $G_1'$ have the same extended $\delta$-folio. Let $G'$ be the graph obtained by replacing $G_1$ with $G_1'$ in the separation $(G_1, G_2)$. Then $G$ and $G'$ have the same extended $\delta$-folio.*

*Proof.* Without loss of generality, we can assume that $R(G) \cap V(G_1) = S$: extending $G_2$ such that $V(G_2)$ fully contains $R(G)$ does not change the statement of the theorem. Under this assumption, it is sufficient to prove the weaker statement that $G$ and $G'$ have the same (not extended) $\delta$-folio (but the condition that $G_1$ and $G_1'$ have the same *extended* $\delta$-folio is not changed). To see this, consider an arbitrary graph $X$ on $R(G)$. Let $X_1$ be the subgraph of $X$ induced by $R(G) \cap V(G_1) = S$ and let $X_2 = X \setminus E(X_1)$. Now $G + X$ has a separation $(G_1 + X_1, G_2 + X_2)$ and $G' + X$ has a separation $(G'_1 + X_1, G'_2 + X_2)$. As $G_1$ and $G'_1$ have the same extended $\delta$-folio, graphs $G_1 + X_1$ and $G'_1 + X_1$ have the same extended $\delta$-folio as well. Therefore, the weaker statement shows that $G + X$ and $G' + X$ have the same $\delta$-folio. As this is true for every $X$ on $R(G)$, it follows that $G$ and $G'$ have the same extended $\delta$-folio.

Let $H$ be a rooted graph with $|E(H)| + \mathrm{is}(H) \leq \delta$ and let $\phi$ be a model of $H$ in $G$. We need to show that $H$ has a model $\phi'$ in $G'$.

We define the graph $X^*$ on $S = R(G) \cap V(G_1)$ such that $uv \in X^*$ for some $u, v \in S$ if there is an edge $e \in E(H)$ such that $\phi(e)$ has a subpath with endpoints $u$ and $v$ and every internal vertex in $V(G_2) \setminus V(G_1)$. For every $uv \in E(X^*)$, let $P_{uv}$ be this subpath. Given a path $P$ in $G$ with endpoints in $V(G_1)$, we denote by $[P]_{G_1}$ the path obtained by replacing subpaths of $P$ that leave $V(G_1)$ by appropriate edges of $X^*$. Similarly, if $Q$ is a path in $G_1 + X^*$, then we denote by $[Q]^G$ the path of $G$ obtained by replacing each edge $uv$ of $X^*$ by the corresponding path $P_{uv}$.

We define a graph $H^*$ and a model $\psi$ of $H^*$ in $G_1 + X^*$ as follows. First, graph $H^*$ contains every vertex $v \in V(H)$ with $\phi(v) \in V(G_1)$; if $v \in R(H)$, then $v$ is in $R(H^*)$ and has the same root number in $H$ and $H^*$. For such vertices, we set $\psi(v) = \phi(v)$. We introduce additional vertices and edges to $H^*$ as follows. We classify each edge $e = uv \in V(H)$ into one of 6 types, and modify $H^*$ accordingly.

(1) $\phi(u), \phi(v) \in V(G_1)$. For each such edge, there is a corresponding edge $e^* = uv$ in $H^*$. We define $\psi(e^*) = [\phi(e)]_{G_1}$.
(2) $\phi(u) \in V(G_1)$, $\phi(v) \notin V(G_1)$, and $\phi(e)$ has an internal vertex in $V(G_1)$. For each such edge, let us introduce a new vertex $v^*_e$ that has the same root number as the last vertex $w$ of $\phi(e)$ (going from $u$ to $v$) that is in $V(G_1)$. Note that this last vertex has to be in $S \subseteq R(G)$, hence it is a root vertex. Let $\psi(v^*_e) = w$. We introduce an edge $e^* = uv^*_e$ in $H^*$ and set $\psi(e^*) = [P]_{G_1}$, where $P$ is the subpath of $\phi(e)$ from $u$ to $w$.
(3) $\phi(u) \in V(G_1)$, $\phi(v) \notin V(G_1)$, and $\phi(e)$ has no internal vertex in $V(G_1)$. This is only possible if $u \in V(G_1) \cap V(G_2)$, hence $u$ is a root. We modify $H^*$ by making $u$ a root (if it is not already a root), having the same root number as $\phi(u)$.
(4) $\phi(u), \phi(v) \notin V(G_1)$, and $\phi(e)$ has no internal vertex in $V(G_1)$. No change is done to $H^*$.
(5) $\phi(u), \phi(v) \notin V(G_1)$, and $\phi(e)$ has a single internal vertex $w$ in $V(G_1)$. This is only possible if $w \in V(G_1) \cap V(G_2)$, and hence $w$ is a root. An isolated root vertex $i^*_e$ is introduced to $H^*$, with the same root number as $w$. Let $\psi(i^*_e) = w$.
(6) $\phi(u), \phi(v) \notin V(G_1)$, and $\phi(e)$ has more than one internal vertex in $V(G_1)$. Let $u_e \neq v_e$ be the first and last vertices, respectively, on $\phi(e)$ (going from $u$ to $v$) that are in $V(G_1)$. Note that $u_e$ and $v_e$ are in $V(G_1) \cap V(G_2)$, hence they are root vertices. Let us introduce root vertices $v^*_e$ and $u^*_e$ in $H^*$ that have the same root numbers as $u_e$ and $v_e$, respectively; let $\psi(u^*_e) = u_e$ and $\psi(v^*_e) = v_e$. Let us also introduce an edge $e^*$ connecting $v^*_e$ and $u^*_e$, and let $\psi(e^*) = [P]_{G_1}$, where $P$ is the subpath of $\phi(e)$ from $u_e$ to $v_e$.

This completes the description of $H^*$. It should be clear that $\psi$ is a model of $H^*$ in $G_1 + X^*$. Furthermore, we claim that $|E(H^*)| + \mathrm{is}(H^*) \leq |E(H)| + \mathrm{is}(H) \leq \delta$. First, for each edge of $H$, we introduce at most one edge in $H^*$ (for type 3–5 edges, we introduce no new edge in $H^*$). Moreover, a vertex of $H^*$ can be isolated only if it was isolated in $H$, or only type 3 edges were adjacent to it, or it was introduced introduced as a vertex $i^*_e$ corresponding to a type 5 edge $e$. This means that the number of isolated vertices in $H^*$ is at most $\mathrm{is}(H)$ plus the number of type 3–5 edges in $H$.

As $H^*$ is a topological minor of $G_1 + X^*$, it is a topological minor of $G'_1 + X^*$ as well; let $\psi'$ be a model of $H^*$ in $G'_1 + X^*$. We show that $\psi'$ can be used to define a model $\phi'$ of $H$ in $G'$, what we need to show. For every $v \in V(H)$ with $\phi(v) \in V(G_1)$, let $\phi'(v) = \psi'(v)$ (as $v \in V(H^*)$ in this case) and for every $v \in V(H)$ with $\phi(v) \in V(G_2) \setminus V(G_1)$, let $\phi'(v) = \phi(v)$. The images of the 6 different type of edges in $H$ are defined as follows.

(1) Let $\phi'(e) := [\psi'(e)]^{G'}$.
(2) Let $w \in S$ be the last vertex on $\phi(e)$ from $u$ to $v$. We obtain $\phi'(e)$ by concatenating $[\psi'(uv^*_e)]^{G'}$ (which goes from $\psi'(u)$ to $w$) and the subpath of $\phi(e)$ from $w$ to $v$.

(3) $\phi'(e) := \phi(e)$.
(4) $\phi'(e) := \phi(e)$.
(5) $\phi'(e) := \phi(e)$.
(6) The path $\phi'(e)$ is obtained by concatenating the subpath of $\phi(e)$ from $u$ to $u_v$, the path $[\psi'(u_e^* v_e^*)]^{G'}$, and the subpath of $\phi(e)$ from $u_v$ to $u$.

It is not difficult to verify that the paths $\phi'(e)$ defined above are internally disjoint. What is important to observe is that if a subpath of $\phi(e)$ is used in the definition above, then every vertex of this subpath in $V(G_1) \cap V(G_2)$ corresponds to a root of $H^*$, hence it cannot conflict with that paths $\psi'(e)$. Thus $\phi'$ is a model of $H$ in $G'$, what we had to show. □

Lemma 2.4 implies that a separation allows us to determine the folio from the folios of two smaller graphs.

**Proposition 2.5.** *Let $(G_1, G_2)$ be a separation of a rooted graph $G$, let $S = V(G_1) \cap V(G_2)$, and suppose that $S \subseteq R(G)$. The extended $\delta$-folio of $G$ can be computed from the extended $\delta$-folios of $G_1$ and $G_2$.*

*Proof.* Let $\mathcal{F}_1$ and $\mathcal{F}_2$ be the extended $\delta$-folios of $G_1$ and $G_2$, respectively. Let use brute force to find minimum representatives $G'_1$ and $G'_2$ of $\mathcal{F}_1$ and $\mathcal{F}_2$, respectively. By definition, we have $|V(G'_1)|, |V(G'_2)| \leq L_{\delta, |R(G)|}$. By Lemma 2.4, replacing $G_1$ with $G'_1$ in the separation $(G_1, G_2)$ does not change the extended $\delta$-folio. With a second application of Lemma 2.4, we can replace $G_2$ with $G'_2$, and obtain a graph $G'$ on at most $2L_{\delta, |R(G)|}$ vertices that have the same extended $\delta$-folio as $G$. The extended $\delta$-folio of $G'$ can be determined by brute force. □

Given a rooted graph $G$, let $w$ be a weight function that assigns a positive integer to each vertex of $V(G)$. The *w-bounded $\delta$-folio* of $G$ contains those members $H$ of the $\delta$-folio of $G$ that have a model $\phi$ satisfying the additional requirement that for every $v \in R(H)$, the degree of $v$ in $H$ is at most $w(\phi(v))$. Note that we do not make any restriction on the degree of a non-root vertex $u$ of $H$, even if $\phi(u)$ happens to be a root vertex of $G$. The term *unbounded $\delta$-folio* is used when we want to emphasize that we are referring to the original definition of $\delta$-folio. The $w$-bounded extended $\delta$-folio is defined analogously. Given a weight function $w$ on the vertices of $G$, we define $w(S) = \sum_{v \in S} w(v)$ for every $S \subseteq V(G)$.

Lemma 2.4 does not remain true for $w$-bounded folios: it is not true that $G$ and $G'$ have the same *w-bounded* extended $\delta$-folio is not sufficient to require that $G_1$ and $G'_1$ have the same *w-bounded* extended $\delta$-folio. The particular point where the proof would fail is that a type 3 edge can make a vertex of $H$ a root which was not a root in $H$, and therefore it is not true that the model $\psi$ is $w$-bounded. However, the proof can be fixed if we impose the additional assumption that $G_1$ and $G'_1$ have the same unbounded extended $(\delta - 1)$-folio. This statement will be used in Section 4 in a situation where the $w$-bounded $\delta$-folio of $G_1$ is easy to determine and we can use recursion to compute the unbounded $(\delta - 1)$-folio.

**Lemma 2.6.** *Let $(G_1, G_2)$ be a separation of a rooted graph $G$, let $S = V(G_1) \cap V(G_2)$, and suppose that $S \subseteq R(G)$. Let $w$ be a weight function that assigns a positive integer to each vertex of $V(G)$. Let $G'_1$ be a rooted graph compatible with $G_1$ such that $G_1$ and $G'_1$ have the same $w$-bounded extended $\delta$-folio and the same unbounded extended $(\delta - 1)$-folio. Let $G'$ be the graph obtained by replacing $G_1$ with $G'_1$ in the separation $(G_1, G_2)$. Then $G$ and $G'$ have the same $w$-bounded extended $\delta$-folio.*

*Proof.* The proof is the same as the proof Lemma 2.4 with one additional argument. Suppose first that $|E(H^*)| + \mathrm{is}(H^*) \leq \delta - 1$. In this case, we know that $H^*$ is in the $(\delta - 1)$-folio of $G'_1 + X^*$ as well, thus the model $\psi'$ exists and the model $\phi'$ can be constructed. Note that $R(G) = R(G_2)$, which means that $\phi'(v) = \phi(v)$ for every root vertex of $H$ and therefore $\phi'$ is $w$-bounded if $\phi$ is $w$-bounded.

Suppose now that $|E(H^*)| + \mathrm{is}(H^*) = \delta$. We claim that in this case $\psi$ is $w$-bounded and hence $H^*$ is in the $w$-bounded $\delta$-folio of $G_1 + X^*$ (not only in the unbounded $\delta$-folio). The vertices in $V(H^*) \setminus V(H)$ have degree at most 1, thus the degree bound holds for such vertices (recall that $w(\psi(v))$ is strictly positive). If a vertex $v \in R(H^*) \cap V(H)$ is in $R(H)$, then $\psi(v) = \phi(v)$ and hence the degree condition holds. Thus we have potential problems only with vertices in $(R(H^*) \cap V(H)) \setminus R(H)$, i.e., vertices that were already present as non-root vertices in $H$, but became roots in $H^*$. The only way such a vertex $u$ could have become a root is if $u$ was incident to a type 3 edge $uv$. If $u$ is isolated in $H^*$, then the degree bound immediately holds. If $u$ is not isolated, then the type 3 edge $uv$ does not create any edge or any new isolated vertex in $H^*$, thus there is at least one edge of $H$ that does not contribute towards $|E(H^*)| + \text{is}(H^*)$, contradicting $|E(H^*)| + \text{is}(H^*) = \delta$. Thus no such vertex $u$ is possible, and it follows that $\psi$ is $w$-bounded. As $G_1$ and $G_1'$ have the same $w$-bounded extended $\delta$-folio, the model $\psi'$ exists, and the rest of the proof is the same as before. □

Analogously to Prop. 2.5, a separation gives a way of determining the $w$-bounded folio.

**Proposition 2.7.** *Let $(G_1, G_2)$ be a separation of a rooted graph $G$, let $S = V(G_1) \cap V(G_2)$, and suppose that $S \subseteq R(G)$. Let $w$ be a weight function that assigns a positive integer to each vertex of $R(G)$. The $w$-bounded extended $\delta$-folio of $G$ can be computed from the $w$-bounded extended $\delta$-folio of $G_1$, the unbounded extended $(\delta - 1)$-folio of $G_1$, and the unbounded extended $\delta$-folio of $G_2$.*

## 3 Algorithmic framework

The main result of the paper is an algorithm FINDFOLIO that determines the extended $\delta$-folio of the given graph.

> FINDFOLIO
> Input: Rooted graph $G$, integer $\delta$.
> Output: The extended $\delta$-folio of $G$.

**Theorem 3.1.** *There is an algorithm satisfying the specification of* FINDFOLIO *that runs in $f_1(\delta, |R(G)|) \cdot |V(G)|^3$ steps, for some computable function $f_1$.*

For technical reasons, we prove Theorem 3.1 in the following form:

**Lemma 3.2.** *There is an algorithm satisfying the specification of* FINDFOLIO *on instances with $|R(G)| \leq 16\delta^2$ that runs in $f_1'(\delta) \cdot |V(G)|^3$ steps, for some computable function $f_1'$.*

It is clear that Lemma 3.2 implies Theorem 3.1: by increasing $\delta$ to, say, $|R(G)|$, the algorithm of Lemma 3.2 can be used even if $|R(G)|$ is arbitrary.

First we design three auxiliary algorithms that either return the extended $\delta$-folio, or some information that is helps our progress: an irrelevant vertex, a clique minor, or an appropriate separation. We say that a set $X$ of vertices is *irrelevant* to the (extended) $\delta$-folio of $G$, if rooted graphs $G$ and $G \setminus X$ have the same (extended) $\delta$-folio. We say that a vertex $v$ is irrelevant if the set $\{v\}$ is irrelevant. Note that even if every vertex of a set $X$ is irrelevant, the set $X$ need not be irrelevant.

> FINDIRRELEVANTORSEPARATION
> Input: Rooted graph $G$, integer $\delta$, integer $L$.
> Output: – The extended $\delta$-folio of $G$, or
> – a vertex $v \in V(G)$ irrelevant to the extended $\delta$-folio of $G$, or
> – a separation $(G_1, G_2)$ of $G$ with $|V(G_1)|, |V(G_2)| \geq L$ and having order at most $4\delta^2$.

We say that $B_1, \ldots, B_k$ are the branch sets of a $K_k$-minor, if they are pairwise disjoint, and for every $1 \leq i < j \leq k$, there is an edge with one endpoint in $B_i$ and one endpoint in $B_j$.

> FINDIRRELEVANTORCLIQUE
> Input: Rooted graph $G$, integer $\delta$, integer $k$.
> Output: – The $\delta$-folio of $G$, or
> – a vertex $v \in V(G)$ irrelevant to the $\delta$-folio of $G$, or
> – the branch sets $B_1, \ldots, B_k$ of a $K_k$-minor in $G$.

> FINDIRRELEVANTORCLIQUEX
> Input: Rooted graph $G$, integer $\delta$, integer $k$.
> Output: – The *extended* $\delta$-folio of $G$, or
> – a vertex $v \in V(G)$ irrelevant to the *extended* $\delta$-folio of $G$, or
> – the branch sets $B_1, \ldots, B_k$ of a $K_k$-minor in $G$.

**Theorem 3.3.** *There is an algorithm satisfying the specification of* FINDIRRELEVANTORCLIQUE *that runs in $f_2(\delta, |R(G)|, k) \cdot |V(G)|$ steps, for some computable function $f_2$.*

**Algorithm 1** FINDFOLIO

1: Let $L := 4\delta^2 + 1$.
2: Let $X := \emptyset$ {$X$ is irrelevant to the extended $\delta$-folio of $G$}
3: Let $Ret = $ FINDIRRELEVANTORSEPARATION$(G \setminus X, \delta, L)$.
4: **if** $Ret$ is the extended $\delta$-folio $\mathcal{F}$ of $G \setminus X$ **then**
5:     **return** $\mathcal{F}$
6: **if** $Ret$ is an irrelevant vertex $v$ **then**
7:     Let $X := X \cup \{v\}$
8:     **goto** 3
9: **if** $Ret$ is a separation $(G_1, G_2)$ of $G \setminus X$ **then**
10:     $S := V(G_1) \cap V(G_2)$
11:     $G_1' := $ AddRoot$(G_1, S)$
12:     $\mathcal{F} = $ FINDFOLIO$(G_1', \delta)$
13:     **if** there is a representative $G_1''$ of $\mathcal{F}$ with at most $L$ vertices **then**
14:         $G'' := (G_1'', G_2)$
15:         $G''' := $ RemoveRoot$(G'', S \setminus R(G))$
16:         return FINDFOLIO$(G''', \delta)$
17:     **else**
18:         Let $L := L + 1$
19:         **goto** 3

Theorem 3.3 is proved in Section 5. It is easy to show that an algorithm for FINDIRRELEVANTORCLIQUE can be used to obtain an algorithm for FINDIRRELEVANTORCLIQUEX:

**Corollary 3.4.** *There is an algorithm satisfying the specification of* FINDIRRELEVANTORCLIQUEX *that runs in* $f_2'(\delta, |R(G)|, k) \cdot |V(G)|$ *steps, for some computable function* $f_2'$.

*Proof.* Let us run the algorithm FINDIRRELEVANTORCLIQUE given by Theorem 3.3 on $G$ with $\delta' := \delta + |R(G)|$ and $k' := k$. If this call returns the $\delta'$-folio of $G$, then by Prop. 2.3(3), we are able to compute and output the extended $\delta$-folio of $G$. If the call returns a vertex $v$ that is irrelevant to the $\delta'$-folio of $G$, then again by Prop. 2.3(3), vertex $v$ is irrelevant to the extended $\delta$-folio of $G$, and hence can be returned as a correct output. Finally, a minor model of a $k$-clique in $G$ is a also a valid output for FINDIRRELEVANTORCLIQUEX. □

Section 4 presents an algorithm for FINDIRRELEVANTORSEPARATION:

**Theorem 3.5.** *There is an algorithm satisfying the specification of* FINDIRRELEVANTORSEPARATION *that runs in* $f_3(\delta, |R(G)|, L) \cdot |V(G)|^2$ *steps, for some computable function* $f_3$.

We prove Theorem 3.5 and Lemma 3.2 by simultaneous induction. In the rest of this section, we prove Lemma 3.2 for some $\delta$, assuming that Theorem 3.5 is true for this $\delta$; while in Section 4, we prove Theorem 3.5 for some $\delta$, assuming that Lemma 3.2 is true for $\delta - 1$. It is clear that these two proofs together prove Theorem 3.5 and Lemma 3.2 for every $\delta \geq 0$.

*Proof (of Lemma 3.2).* Let $L^* = \max\{L_{\delta, 12\delta^2}, 16\delta^2\}$. This constant will be required only for the analysis of the algorithm and it does not appear explictly in the description of the algorithm. Algorithm 1 shows the algorithm in pseudocode. The functions AddRoot$(G, S)$ and RemoveRoot$(G, S)$ return a rooted graph where $S$ is added to/removed from the set of roots, respectively.

Let $L := 4\delta^2 + 1$. We will increase $L$ during the algorithm, but (as we shall see) $L \leq L^*$ will always hold. Initially we set $X := \emptyset$; it will always hold that the set of vertices $X$ is irrelevant to the extended $\delta$-folio of $G$.

Let us run algorithm FINDIRRELEVANTORSEPARATION of Theorem 3.5 with $G \setminus X$, $\delta$, and $L$. If the output is the extended $\delta$-folio of $G \setminus X$, then we are done. If the output is a vertex $v$ irrelevant to the extended $\delta$-folio of $G \setminus X$, then let $X := X \cup \{v\}$ and call FINDIRRELEVANTORSEPARATION again. It is clear that the new $X$ is irrelevant to the extended $\delta$-folio of $G$. Suppose that (after returning some number of irrelevant vertices) FINDIRRELEVANTORSEPARATION returns a separation $(G_1, G_2)$ of $G \setminus X$ with $|V(G_1)|, |V(G_2)| \geq L$ and having order at most $4\delta^2$. Note that $L > 4\delta^2$, and hence $|V(G_1) \setminus V(G_2)|, |V(G_2) \setminus V(G_1)| > 0$.

Let $G'$, $G'_1$, $G'_2$ be the same as $G \setminus X$, $G_1$, and $G_2$, respectively, with the difference that every vertex of $S = V(G_1) \cap V(G_2)$ is a root (in addition to the original roots). Without loss of generality, we can assume that $|R(G_1)| \leq |R(G_2)|$ and hence $|R(G'_1)| \leq |R(G)|/2 + |S| \leq 12\delta^2$. Let us call FINDFOLIO recursively to find the extended $\delta$-folio of $G'_1$ and then let us try to construct by brute force a representative $G''_1$ of this folio having at most $L$ vertices. If we do not find such a representative, then we increase $L$ by one, and go back to calling FINDIRRELEVANTORSEPARATION (note that this is possible only if $L < L_{\delta, 12\delta^2} \leq L^*$, thus we never increase $L$ above $L^*$). Otherwise, we replace $G'_1$ with $G''_1$ in the separation $(G'_1, G'_2)$; let $G''$ be the new graph. By Lemma 2.4, $G'$ and $G''$ have the same extended $\delta$-folio. Let $G'''$ be the graph obtained from $G''$ by making those vertices of $S$ non-roots that are non-roots in $G$ (i.e., $|R(G''')| = |R(G)|$). It is clear that the extended $\delta$-folio of $G \setminus X$ and $G'''$ are the same. Thus we can finish the algorithm by recursively calling FINDFOLIO on $G'''$ (note that $|R(G''')| \leq 16\delta^2$).

It is obvious from the description that the answer returned by the algorithm is correct. Note that $|V(G'_1)|, |V(G''')| < |V(G)|$, thus this recursive procedure always terminates.

We need to show that the number of steps can be bounded by $g(\delta) \cdot |V(G)|^3$ for some function $g$. The running time required for instances with at most $L^* + 1$ vertices can be bounded by a constant depending only on $\delta$. We show that there is a function $g'$ such that the running time can be bounded by $g'(\delta)(|V(G) - L^* - 1|)|V(G)|^2$ for instances with $|V(G)| > L^* + 1$. We prove by induction on $|V(G)|$ that this holds if $g'(\delta)$ is sufficiently large.

Let us bound first the number of steps without the calls to FINDIRRELEVANTORSEPARATION and the recursive calls to FINDFOLIO. Let $x$ be the number of times FINDIRRELEVANTORSEPARATION returned an irrelevant vertex. Then FINDIRRELEVANTORSEPARATION was called at most $x + L^*$ times (each call either returned an irrelevant vertex or increased $L$, but $L \leq L^*$ always hold). Therefore, each line is executed at most $x + L^*$ times. Each step can be done in linear time in the size of the graph, thus we can bound the running time by $c_1 \cdot (x+1)|V(G)|^2$ for some constant $c_1$ depending on $\delta$. By Theorem 3.5, each call to FINDIRRELEVANTORSEPARATION can be bounded by $f_3(\delta, 16\delta^2, L)|V(G)|^2$ steps and the maximum possible value of $L$ is a function of $\delta$, thus the total time required for these calls can be bounded by $c_2 \cdot (x+1)|V(G)|^2$ for some constant $c_2$ depending only on $\delta$.

Finally, let us bound the running time of the recursive calls to FINDFOLIO. If $|V(G'_1)| \leq L^* + 1$ or $|V(G''')| \leq L^* + 1$, then the number of steps of these calls can be bounded by a constant depending only on $\delta$. Let us assume in the following that $|V(G'_1)|, |V(G''')| > L^* + 1$. As we noted earlier, $|V(G'_1)|, |V(G''')| < |V(G)|$, thus the induction hypothesis can be used to bound the running time of these calls. Therefore, the total running time can be bounded as follows:

$$(c_1 + c_2)(x+1)|V(G)|^2 + g'(\delta)(|V(G'_1)| - L^* - 1)|V(G'_1)|^2 + g'(\delta)(|V(G''')| - L^* - 1)|V(G''')|^2$$
$$\leq g'(\delta)\big((x+1) + |V(G'_1)| - L^* - 1 + |V(G''')| - L^* - 1\big)|V(G)|^2$$
$$\leq g'(\delta)\big((x+1) + |V(G'_1)| - L^* - 1 + |V(G'_2) \setminus V(G'_1)| - 1\big)|V(G)|^2$$
$$\leq g'(\delta)(|V(G)| - L^* - 1)|V(G)|^2.$$

In the first inequality, we assume that $g'(\delta) \geq c_1 + c_2$. The second inequality follows from $|V(G''')| = |V(G'_1) \cup V(G'_2)|$ and $|V(G''_1)| \leq L \leq L^*$. The last inequality follows from $|X| + |V(G'_1) \cup V(G'_2)| = |V(G)|$. □

## 4 Using a large clique minor

In this section, we prove Theorem 3.5 for some $\delta$, assuming that Lemma 3.2 holds for $\delta - 1$. We use the following lemma due to Robertson and Seymour ((5.4) of [11]):

**Lemma 4.1.** *Let $G$ be a graph and $Z \subseteq V(G)$. Let $t \geq (3/2) \cdot |Z|$, and let $B_1, \ldots, B_k \subseteq V(G)$ be the branch sets of a $K_k$-minor of $G$. Suppose that there is no separation $(G_1, G_2)$ of $G$ of order $< |Z|$ with $Z \subseteq V(G_1)$ and $B_b \cap V(G_1) = \emptyset$ for some $b \in [k]$. Then for every partition $(Z_1, \ldots, Z_n)$ of $Z$ into nonempty subsets there are pairwise disjoint connected subgraphs $T_1, \ldots, T_n \subseteq G$ such that $V(T_i) \cap Z = Z_i$ for all $i \in [n]$.*

We say that the $\delta$-folio of a graph is *generic* if it is as large as possible: it contains every rooted graph $H$ with $E(H) + \text{is}(H) \leq \delta$ and $\rho_H(R(H)) \subseteq \rho_G(R(G))$. We say that the $\delta$-folio of a graph is *rooted-generic* if it contains every such graph $H$ with the additional condition that every vertex of $H$ is rooted (thus generic implies rooted-generic, but not necessarily the other way). The notions of generic and rooted-generic are defined analogously for the extended and $w$-bounded folios. Note that if $G$ has a generic $\delta$-folio, then $G + X$ has generic $\delta$-folio for any graph $X$ on $R(G)$: adding edges can only add more graphs to the folio. Thus the extended $\delta$-folio of $G$ is generic if and only if the $\delta$-folio is generic. We can use Lemma 4.1 to obtain sufficient conditions for generic folios:

**Lemma 4.2.** *Let $G$ be a rooted graph. Let $w$ be a positive integer weight function on $V(G)$. Let $k \geq (3/2) \cdot w(R(G))$, and let $B_1, \ldots, B_k \subseteq V(G)$ be the branch sets of a $K_k$-minor of $G$. Suppose that there is no separation $(G_1, G_2)$ of $G$ with $w(V(G_1) \cap V(G_2)) < w(R(G))$, $R(G) \subseteq V(G_1)$, and $B_i \cap V(G_1) = \emptyset$ for some $i \in [k]$.*

*(1) The $w$-bounded $\delta$-folio of $G$ is rooted-generic.*
*(2) If there are at least $2\delta$ vertices $v$ in $R(G)$ with $w(v) \geq 2\delta$, then the $w$-bounded $\delta$-folio of $G$ is generic.*

*Proof.* We need to show that every possible candidate $H$ is in the $w$-bounded $\delta$-folio of $G$. Suppose therefore that $H$ is a rooted graph with $|E(H)| + \text{is}(H) \leq \delta$, $R(H) = V(H)$, and $\rho_H(R(H)) \subseteq \rho_G(R(G))$. For every $u \in V(H)$, let $\phi(u)$ be the vertex of $G$ with the same root number as $u$ and assume that $d_H(u) \leq w(\phi(u))$ for every $u \in V(H)$. We need to show that $H$ is a topological minor of $G$, i.e., $\phi$ can be extended to a model of $H$ in $G$.

For every $v \in V(G)$, let us define $w'(v) = d_H(u)$ if $v = \phi(u)$ for some $u \in V(H)$, and let $w'(v) = w(v)$ if there is no such $u$. Clearly, $w'(v) \leq w(v)$ for every $v \in V(G)$: the degree condition holds for every $v \in R(H) = V(H)$ in $\phi$. Let $G'$ be the graph obtained from $G$ by extending each vertex $z \in R(G)$ into a clique $K_z$ of size $w'(z)$, i.e., we introduce $w'(z) - 1$ new vertices that are adjacent to each other, to vertex $z$, and to every neighbor of $z$. The clique $K_z$ contains $z$ and these $w'(z) - 1$ new vertices. Let $Z := \bigcup_{z \in R(G)} K_z$. Let us show first that the conditions of Lemma 4.1 hold for $Z$ in $G'$. Suppose for contradiction that $(G'_1, G'_2)$ is a separation of $G'$ of order less than $|Z| = w'(R(G)) \leq w(R(G))$ with $Z \subseteq V(G'_1)$ and $B_b \subseteq V(G'_2) \setminus V(G'_1)$ for some $b \in [k]$. Let $S' := V(G'_1) \cap V(G'_2)$ be the separator. Without loss of generality, we may assume that for all $z \in R(G)$, either $K_z \cap S' = \emptyset$ or $K_z \subseteq S'$. Let $G_1 := G'_1 \setminus (Z \setminus R(G))$ and $G_2 := G'_2 \setminus (Z \setminus R(G))$. Then $(G_1, G_2)$ is a separation of $G$; let $S = V(G_1) \cap V(G_2)$ be the separator. Now it is clear that $w(S) = |S'| < |Z| = w'(R(G)) \leq w(R(G))$. However, we also have $R(G) \subseteq V(G_1)$ and $B_b \cap V(G_1) = \emptyset$, contradicting the assumption of the lemma being proved. Thus we can conclude that there is no such separation $(G'_1, G'_2)$, and the conditions of Lemma 4.1 hold for $Z'$ and $G'$.

Let us partition $Z'$ in such a way that for every edge $uv \in E(H)$, there is a 2-element class of the partition consisting of a vertex in $K_{\phi(u)}$ and a vertex in $K_{\phi(v)}$. As $K_{\phi(u)}$ and $K_{\phi(v)}$ contain exactly $d_H(u)$ and $d_H(v)$ vertices, respectively, such a partition exists. Lemma 4.1 gives a set of pairwise disjoint subgraphs, one for each class of the partition. For every edge $uv \in E(H)$, let us denote by $T_{uv}$ the connected subgraph corresponding to the class consisting of a vertex of $K_{\phi(u)}$ and a vertex of $K_{\phi(v)}$, and let us chose a path $P'_{uv}$ in $T_{uv}$ that goes from a vertex of $K_{\phi(u)}$ to a vertex of $K_{\phi(v)}$. It is clear that the collection $\mathcal{P}'$ of $|E(H)|$ paths obtained this way are pairwise disjoint in $G'$. Let us define $P_{uv}$ such that whenever $P'_{uv}$ contains a vertex of some $K_z$, then we replace it by $z$; let $\mathcal{P}$ be the collection of these paths $P_{uv}$ for every $uv \in E(H)$. Observe that the way $G'$ was defined ensures that $P_{uv}$ is a path in $G$. We claim that the paths in $\mathcal{P}$ are pairwise internally disjoint in $G$. As the paths in $\mathcal{P}'$ are pairwise disjoint, the only possible problem is that for some $w \in V(H)$, vertex $\phi(w)$ is an internal vertex of some path $P_{uv}$ with $w \notin \{u, v\}$. However, there are $d_H(w) = |K_{\phi(w)}|$ paths in $\mathcal{P}$ whose endpoint is $\phi(w)$ and hence the disjointness $\mathcal{P}'$ ensure that there cannot be more than $d_H(w)$ paths using vertex $\phi(w)$. We finish the proof of the first statement by extending $\phi$ into a model of $H$ by defining $\phi(uv)$ to be the path $P_{uv}$.

To prove the second statement, let $H$ be a rooted graph with $|E(H)| + \text{is}(H) \leq \delta$. Let us obtain $H'$ by making every vertex of $H'$ a root: if $v \in V(H)$ is not rooted, then let us assign to it a root number that appears on a vertex $v \in R(G)$ with $w(v) \geq \delta$ and is not already used by a vertex of $H$. As $|V(H)| \leq 2\delta$, the conditions of the lemma show that we can assign root numbers this way. Since the $w$-bounded $\delta$-folio of $G$ is rooted-generic, $H'$ is topological minor of $G$, which means that $H$ is also a topological minor of $G$. $\square$

We prove Theorem 3.5, under the assumption that Theorem 3.1 is true for $\delta - 1$. Let us define the following

constants:

$$h := 2\delta$$
$$s := 4\delta^2$$
$$k := \max\{L, 10\delta^2\} + |R(G)|$$

One possible correct output of FINDIRRELEVANTORSEPARATION is a separation $(G_1, G_2)$ of $G$ with $|V(G_1)|, |V(G_2)| \geq L$ and $|V(G_1) \cap V(G_2)| \leq s$. We refer to this as *finding a small separator.*

The algorithm for FINDIRRELEVANTORSEPARATION starts by calling FINDIRRELEVANTORCLIQUEX for $G$, $\delta$, and $k$. If FINDIRRELEVANTORSEPARATION returns an irrelevant vertex or the extended $\delta$-folio of $G$, then this is a valid output for FINDIRRELEVANTORSEPARATION as well. Suppose therefore that FINDIRRELEVANTORSEPARATION returns a $k$-clique minor with branch sets $B_1, \ldots, B_k$. As at most $|R(G)|$ of these sets intersect $R(G)$, we can assume without loss of generality that $B_1, \ldots, B_L$ are disjoint from $R(G)$.

The rest of the section discusses two cases depending on the number of vertices with degree at least $L$ in $G'$.

### 4.1 Case 1: Many high-degree vertices

Suppose that there are at least $h$ vertices with degree at least $L$. Let $U$ be a set of $h$ such vertices.

Let us enumerate every nonempty subset of size at most $2\delta$ of $|R(G)|$; let $R_1, \ldots, R_t$ be these subsets. Let $w_i$ be a weight assignment on $V(G)$ such that $w(v) = \delta$ if $v \in R_i \cup U$ and $w(v) = 1$ otherwise. By Proposition 3, the folio of $G$ can be obtained from the folios $G$ with respect to $R_1, \ldots, R_t$. Furthermore, the $w_i$-bounded $\delta$-folio of $G$ with respect to $R_i$ is obviously the same as the unbounded $\delta$-folio with respect to $R_i$.

For every $1 \leq i \leq t$, we compute a separation $(G_1^i, G_2^i)$ of $G$ such that $R_i \cup U \subseteq V(G_1^i)$, there is a $1 \leq b \leq L$ with $B_b \subseteq V(G_2^i) \setminus V(G_1^i)$, and $w_i(V(G_1^i) \cap V(G_2^i))$ is as small as possible. Such a separation $(G_1^i, G_2^i)$ can be done by running, for every $1 \leq j \leq L$, a weighted minimum vertex cutset algorithm to find a set of vertices that separates $R_i \cup U$ and $B_j$; among these $L$ separations, we define $(G_1^i, G_2^i)$ to be the one that minimizes $w_i(V(G_1^i) \cap V(G_2^i))$. Let $S_i := V(G_1^i) \cap V(G_2^i)$.

Note that $(G[R_i \cup U], G \setminus E(G[R_i \cup U]))$ is always a separation that satisfies the requirements, thus we can assume that $w_i(S_i) \leq w(R_i \cup U)\delta(2\delta + h) = s$. As each of $B_1, \ldots, B_L$ intersects $V(G_2^i)$, we have $|V(G_2^i)| \geq L$. This means that if $|V(G_1^i)| \geq L$ also holds, then separation $(G_1^i, G_2^i)$ is a small separation that can be returned as a valid output of FINDIRRELEVANTORSEPARATION. Thus we can assume in the following that $|V(G_1^i)| < L$. This implies that $U \subseteq S_i$: if some $u \in U$ is not in $S_i$, then every neighbor of $u$ is in $V(G_1^i)$, and $|V(G_1^i)| \geq L$ follows.

We use Lemma 4.2 to show that the $w_i$-bounded $\delta$-folio of $G_2^i$ is generic *with respect to $S_i$*. At most $|S_i| \leq w_i(R_i \cup U) \leq \delta(|R_i| + |U|) \leq 4\delta^2$ of the sets $B_1, \ldots, B_L$ intersect $S_i$, thus we can suppose without loss of generality that $B_1, \ldots, B_{6\delta^2}$ are disjoint from $S_i$. Suppose that $G_2^i$ has a separation $(F_1, F_2)$ contradicting the conditions of Lemma 4.2: $S_i \subseteq V(F_1)$, $B_b \subseteq V(F_2) \setminus V(F_1)$ for some $1 \leq b \leq 6\delta^2$, and $w_i(V(F_1) \cap V(F_2)) < w_i(S_i)$. Such a separation can be extended to a separation $(F_1', F_2')$ of $G$ with $V(G_1^i) \subseteq V(F_1')$, $V(F_1') \cap V(F_2) = V(F_1) \cap V(F_2)$ and $B_b \subseteq V(F_2') \setminus V(F_1')$. However, such a separation would contradict the minimality of the choice of $S_i$. Thus the conditions of Lemma 4.2 hold, and the $w_i$-bounded $\delta$-folio of $G_2^i$ is generic with respect to $S_i$.

We use Proposition 2.7 to compute the $w_i$-bounded $\delta$-folio of $G$ with respect to $R_i \cup S_i$; by Proposition 2.3(2), this can be used to compute the $w_i$-bounded $\delta$-folio of $G$ with respect to $R_i$. As $|V(G_1^i)| < L$, the extended $\delta$-folio of $G_1^i$ with respect to $R_i \cup S_i$ can be determined by brute force in time depending only on $L$. We can determine the (unbounded) extended $(\delta - 1)$-folio of $G_2^i$ with respect to $S_i$ by calling FINDFOLIO (recall that we assume in this section that Lemma 3.2 holds for $\delta - 1$ and $|S_i| \leq 4\delta^2 \leq 16(\delta - 1)^2$, satisfying the conditions of Lemma 3.2). We have shown above that the extended $w_i$-bounded $\delta$-folio of $G_2^i$ with respect to $S_i$ is generic. Thus we have all the information required by Prop. 2.7 at our disposal to compute the $w_i$-bounded $\delta$-folio of $G$ with respect to $R_i \cup S_i$.

## 4.2 Case 2: Few high-degree vertices

Let $U$ be the set of all vertices in $G'$ with degree at least $L$; we suppose in this case that $|U| < h$. To determine the extended $\delta$-folio of $G$, for every graph $X$ on $R(G)$, we need to determine the $\delta$-folio of $G+X$. Fixing such an $X$, we set $G' = G + X$ and proceed the following way.

We define a graph $F$ on vertex set $V(G') \setminus U$, where two vertices are adjacent if their distance in $G' \setminus U$ is at most $2L$. As the maximum degree of $G' \setminus U$ is at most $L' = L + |R(G)|$, the maximum degree of $F$ is at most $(L')^{2L'+1}$. We say that a subset $C \subseteq V(G') \setminus U$ of vertices is a *cluster* if $F[C]$ is connected. Observe that the maximum number of clusters of size at most $x$ that contain a vertex $v \in V(G') \setminus U$ can be bounded by a function of the maximum degree of $F$ and $x$. Therefore, assuming $\delta$, $|R(G)|$, and $L$ are fixed constants, the total number of clusters of size at most $2\delta$ is linear in $|V(G')|$. Let $C_1, \dots, C_t$ be an enumeration of the clusters of size at most $2\delta$.

For every $1 \le i \le t$, let $w_i$ be a weight function on $V(G') \setminus U$ defined as $w_i(v) = \delta$ for $v \in C_i$ and $w_i(v) = 1$ otherwise. For every $1 \le i \le t$, let us choose a separation $(G_1^i, G_2^i)$ of $G' \setminus U$ such that $C_i \subseteq V(G_1^i)$, there is a branch set $B_b$ with $B_b \subseteq V(G_2^i) \setminus V(G_1^i)$, and $w_i(V(G_1^i) \cap V(G_2^i))$ is minimum possible. It is easy to see that we can choose the separation such that every connected component of $G_1^i$ contains a vertex of $C_i$. Let $D_i = V(G_1^i)$ and $S_i = V(G_1^i) \cap V(G_2^i)$. The separation $(G'[C_i], G' \setminus E(G'[C_i]))$ and the minimality of $w_i(S_i)$ shows that $w_i(S_i) \le w_i(C_i) \le 2\delta \cdot \delta$ and hence $|S_i| \le w_i(S_i) \le 2\delta^2$. Every branch set of the clique intersects $V(G_2^i)$, which means that $|V(G_2^i)| \ge L$. If $|V(G_1^i)| \ge L$ also holds, then $G'$ has a small separation $(G_1, G_2)$ with $V(G_1) = V(G_1^i) \cup u$, $V(G_2) = V(G_2^i) \cup U$, and $|V(G_1) \cap V(G_2)| = |V(G_1^i) \cap V(G_2^i)| + |U| \le s$, which we can return. Thus in the following, we can assume that $|D_i| < L$.

We say that two clusters $C_{i_1}$ and $C_{i_2}$ are *independent* if there is no edge between $C_{i_1}$ and $C_{i_2}$ in $F$.

**Proposition 4.3.** *If clusters $C_{i_1}$ and $C_{i_2}$ are independent, then $D_{i_1} \cap D_{i_2} = \emptyset$.*

*Proof.* Let us choose a vertex $v \in D_{i_1} \cap D_{i_2}$. As $|D_{i_1}| < L$ and the component of $G_1^{i_1}$ containing $v$ contains a vertex of $C_{i_1}$, vertex $v$ is at distance at most $L$ from some vertex of $C_{i_1}$ in $G_1^{i_1}$, and therefore in $G' \setminus U$. Similarly, $v$ is at distance at most $L$ from some vertex of $C_{i_2}$ in $G' \setminus U$. Thus there is an edge in $F$ between a vertex of $C_{i_1}$ and a vertex of $C_{i_2}$, a contradiction. □

**Definition 4.4.** We say that clusters $C_{i_1}$ and $C_{i_2}$ are *equivalent* if there is a rooted isomorphism between the graphs $G[D_{i_1} \cup U]$ and $G[D_{i_2} \cup U]$ that is the identity on $U$, maps $S_{i_1}$ to $S_{i_2}$, and maps $C_{i_1}$ to $C_{i_2}$.

The following proposition is easy to prove:

**Proposition 4.5.** *The number of equivalence classes of the clusters can be bounded by a function of $\delta$ and $L$.*

As we shall see, the topological minor is realized by a small number of clusters and paths connecting them. The following definition tries to capture which paths are inside a cluster and which paths are between clusters.

**Definition 4.6.** Let $H$ be a rooted graph. A *scheme* of $H$ is a pair $(H', H'_*)$ of rooted graphs, where

(1) $H'$ is a subdivision of $H$ (the new vertices are not roots),
(2) $H'_*$ is a subgraph of $H'$, and
(3) every vertex of $V(H'_*) \setminus V(H)$ has degree at most 1 in $H'_*$.

For every $r$-tuple $\mathcal{C} = (C_{i_1}, \dots, C_{i_r})$ of clusters, we define $C^{\mathcal{C}} = \bigcup_{i=1}^r C_{i_j}$, $D^{\mathcal{C}} = \bigcup_{i=1}^r D_{i_j}$, and $S^{\mathcal{C}} = \bigcup_{i=1}^r S_{i_j}$. We define two graphs: $G_1^{\mathcal{C}} = G'[U \cup D^{\mathcal{C}}]$ and $G_2^{\mathcal{C}} = G' \setminus (D^{\mathcal{C}} \setminus S^{\mathcal{C}})$. Note that $(G_1^{\mathcal{C}}, G_2^{\mathcal{C}})$ is a separation of $G$. We also define a weight function $w^{\mathcal{C}}$ on $V(G)$ that is $\delta$ on every vertex of $U \cup C^{\mathcal{C}}$ and $1$ on every other vertex.

**Definition 4.7.** Let $H$ be a rooted graph and let $(H', H'_*)$ be a scheme of $H$. Let $\mathcal{C} = (C_{i_1}, \dots, C_{i_r})$ be an $r$-tuple of clusters. We say that this tuple *realizes* the scheme $(H', H'_*)$ if $H' \setminus E(H'_*)$ has a model $\phi$ in $G_1^{\mathcal{C}}$ such that

(1) every vertex of $V(H)$ is mapped to $U \cup C^{\mathcal{C}}$,
(2) every vertex of $V(H'_*)$ is mapped to $U \cup S^{\mathcal{C}}$, and
(3) for every $e \in E(H') \setminus E(H'_*)$, the internal vertices of $\phi(e)$ are not in $U \cup S^{\mathcal{C}}$.

Roughly speaking, what we want to show is that $H$ is a topological minor of $G$ if and only if there is a tuple of independent clusters that realizes a scheme of $H$ (Lemmas 4.8 and 4.10). Therefore, deciding whether $H$ is a topological minor essentially reduces to finding a tuple of independent clusters that realize a given scheme of $H$. As the clusters can be classified into a bounded number of equivalence classes, the main difficulty is to find independent clusters of given types, which can be solved using standard techniques.

We first prove that if a rooted graph $H$ has model in $G'$, then $H$ has a scheme that realized by some tuple of clusters. We hope the proof sheds light on why schemes are defined this way.

**Lemma 4.8.** *Let $H$ be a rooted graph in the $\delta$-folio of $G'$. Then there is a scheme $(H', H'_*)$ of $H$ with $|V(H')| \leq 4\delta + 2\delta^2$ and a tuple $\mathcal{C} = (C_{i_1}, \ldots, C_{i_r})$ of pairwise independent clusters with $r \leq 2\delta$ that realizes $(H', H'_*)$.*

*Proof.* Let $\phi$ be a model of $H$ in $G'$. Let $C = \{\phi(v) \mid v \in V(H)\} \setminus U$. Each connected component of $F[C]$ is a cluster; let $\mathcal{C} = (C_{i_1}, \ldots, C_{i_r})$ be these connected components. Clearly, these clusters are pairwise independent and $r \leq |V(H)| \leq 2\delta$. Due to a minor technical detail, we need to handle some vertices of $S^\mathcal{C} \cup U$ in a special way. We define $X$ to contain a vertex $v \in S^\mathcal{C} \cup U$ if $v$ is an internal vertex of $\phi(e)$ for some $e \in E(H)$ and both neighbors of $v$ in $\phi(e)$ are in $V(G_2^\mathcal{C})$.

If for some $e \in V(H)$, path $\phi(e)$ contains $m$ internal vertices in $(S^\mathcal{C} \cup U) \setminus X$, then let us subdivide $e$ with $m$ new (non-root) vertices; let $H'$ be the rooted graph obtained this way. As $|S^\mathcal{C} \cup U| \leq 2\delta^2 + 2\delta$, we have $|V(H')| \leq 4\delta + 2\delta^2$. The model $\phi$ gives a model $\phi'$ of $H$ in $G$ the obvious way (every new vertex of the subdivision is mapped to a vertex in $(S^\mathcal{C} \cup U) \setminus X$). Let $H'_*$ be the subgraph of $H'$ that contains those vertices $v$ for which $\phi'(v) \in (S^\mathcal{C} \cup U) \setminus X$ and those edges $e$ for which $\phi'(e)$ is fully contained in $G_2^\mathcal{C}$.

We claim that $(H', H'_*)$ is a scheme of $H$ and $\mathcal{C}$ realizes this scheme. Conditions 1 and 2 of Definition 4.6 are easy to verify. To check condition 3, suppose that vertex $v \in V(H'_*) \setminus V(H)$ has degree more than 1. Since vertex $v$ was obtained as the subdivision of an edge $e \in E(H)$, vertex $v$ has degree exactly 2 in $H'_*$ and $\phi'(v) \in (S^\mathcal{C} \cup U) \setminus X$. Let $e_1$ and $e_2$ be the two edges incident to $v$ in $H'_*$. By definition of $H'_*$, $\phi'(e_1)$ and $\phi'(e_2)$ are fully contained in $G_2^\mathcal{C}$. Thus the two neighbors of $\phi(v)$ in $\phi(e)$ are both in $V(G_2^\mathcal{C})$, implying that $\phi(v) \in X$, a contradiction.

Finally, we show that $\phi'$ defines a model of $H' \setminus E(H'_*)$ in $G_1^\mathcal{C}$ satisfying the conditions of Definition 4.7. Let us verify that the images of the vertices and edges are indeed in $G_1^\mathcal{C}$. It is clear that $\phi'(v) \in V(G_1^\mathcal{C})$ for every $v \in V(H')$. Let us prove that $\phi'(e)$ is fully contained in $V(G_1^\mathcal{C})$ for every $e \in E(H') \setminus E(H'_*)$. In fact, we show that $\phi'(e)$ has no internal vertex in $V(G_2^\mathcal{C})$. Suppose that $\phi'(e)$ has an internal vertex $u_2 \in V(G_2^\mathcal{C})$. As $e \notin E(H'_*)$, path $\phi'(e)$ contains a vertex $u_1 \in V(G_1^\mathcal{C}) \setminus V(G_2^\mathcal{C})$ ($u_1$ can be an endpoint of $\phi'(e)$). Going from $u_1$ to $u_2$ on $\phi'(e)$, let $u$ be the first vertex of $V(G_2^\mathcal{C})$; clearly, $u \in S \cup U$ and $u \neq u_1$. Now $u$ is an internal vertex of $\phi'(e)$, and the vertex preceding $u$ is not in $V(G_2^\mathcal{C})$. Thus $u \in (S^\mathcal{C} \cup U) \setminus X$, which means that $u$ should be the image of a vertex of $H'$ in $\phi'$, a contradiction. Therefore, $\phi'(e)$ has no internal vertex in $V(G_2^\mathcal{C})$ and in particular $\phi'(e)$ is fully contained in $V(G_1^\mathcal{C})$ for every $e \in E(H') \setminus E(H'_*)$. This means that $\phi'$ is indeed a model of $H' \setminus E(H'_*)$ in $G_1^\mathcal{C}$ and we also verified condition 3 of Definition 4.7. Conditions 1 and 2 are straightforward to check. □

We prove now the converse of Lemma 4.8. We show first that the $w_i$-bounded folio of $G_2^\mathcal{C}$ is rooted-generic (Lemma 4.9). Then we use this fact to route the edges of $H'_*$ when constructing a model of $H'$ in $G'$ (Lemma 4.10).

**Lemma 4.9.** *Let $\mathcal{C} = (C_{i_1}, \ldots, C_{i_r})$ be a tuple of pairwise independent clusters. Either the $w^\mathcal{C}$-bounded $w^\mathcal{C}(S^\mathcal{C})$-folio of $G_2^\mathcal{C}$ with respect to $U \cup S^\mathcal{C}$ is rooted-generic (and we can find a model of every graph in the folio), or we can find a separation $(G'_1, G'_2)$ of $G'$ with $|V(G'_1)|, |V(G'_2)| \geq L$ and $|V(G'_1) \cap V(G'_2)| \leq s$.*

*Proof.* If the conditions of Lemma 4.2 hold for $G_2^\mathcal{C}$, $w^\mathcal{C}$, and set of roots $U \cup S^\mathcal{C}$, then we are done. Suppose therefore that there is a separation $(F_1, F_2)$ of $G_2^\mathcal{C}$ violating the conditions of Lemma 4.2. There is a corresponding separation $(G'_1, G'_2)$ of $G'$ with $V(F_1) \cap V(F_2) = V(G'_1) \cap V(G'_2)$, $V(G'_1) \subseteq V(F_1)$, and $V(G'_2) = V(F_2)$. Let $S' = V(F_1) \cap V(F_2) = V(G'_1) \cap V(G'_2)$, it is clear that $|S'| \leq w^\mathcal{C}(U \cup S^\mathcal{C}) \leq s$. As $B_b \subseteq V(G'_2)$, we also have $|V(G'_2)| \geq L$. If $|V(G'_1)| \geq L$, then we can return the small separation $(G'_1, G'_2)$. Thus in the following, we can assume that $|V(G'_1)| \leq L$. In particular, this means that $U \subseteq S'$: if $u \in V(G'_1) \setminus V(G'_2)$ for some $u \in U$, then every neighbor of $u$ is in $V(G'_1)$ and $|V(G'_1)| \geq L$ follows.

Let $S'_{i_j}$ be the set of those vertices of $S' \setminus U$ that can be reached from $S_{i_j} \subseteq V(G'_1) \setminus U$ by a path in $G'_1 \setminus U$. We claim that these sets are pairwise disjoint for $j = 1, \ldots, r$. Suppose without loss of generality that there is a vertex $v \in S'_{i_1} \cap S'_{i_2}$. This means that there is a vertex $v_1 \in S_{i_1}$ and a vertex $v_2 \in S_{i_2}$ that are in the same connected component $K$ of $G'_1 \setminus U$ as $v$. Note that $D_{i_1}$ and $D_{i_2}$ are fully contained in $G'_1 \setminus U$, thus there is a vertex $c_1 \in C_{i_1}$ and a vertex $c_2 \in C_{i_2}$ in this connected component $K$. As clusters $C_{i_1}$ and $C_{i_2}$ are independent by assumption, the distance of $c_1$ and $c_2$ is at least $2L$ in $G'_1 \setminus U$, which means that $|V(G'_1)| \geq 2L$, a contradiction.

As $U, S'_{i_1}, \ldots, S'_{i_r}$ are pairwise disjoint and $U \subseteq S$, the only way $w^{\mathcal{C}}(S') < w^{\mathcal{C}}(S)$ is only possible if $w^{\mathcal{C}}(S'_{i_j}) < w^{\mathcal{C}}(S_{i_j})$ for some $1 \leq j \leq r$. However, in this case there is a separation $(G_1^{i_j}, G_2^{i_j})$ of $G' \setminus U$ with $V(G_1^{i_j}) \cap V(G_2^{i_j}) = S'_{i_j}$, $D_{i_j} \subseteq V(G_1^{i_j})$, and $B_b \subseteq V(G_2^{i_j}) \setminus V(G_1^{i_j})$ for some branch set $B_b$. This contradicts the minimality of the choice of $S_{i_j}$. □

**Lemma 4.10.** *Let $H$ be a rooted graph and $(H', H'_*)$ be a scheme of $H$. Let $\mathcal{C} = (C_{i_1}, \ldots, C_{i_r})$ be an r-tuple of pairwise independent clusters that realizes $(H', H'_*)$. Then we can find either a model of $H$ in $G'$ or a separation $(G'_1, G'_2)$ of $G'$ with $|V(G'_1)|, |V(G'_2)| \geq L$ and $|V(G'_1) \cap V(G'_2)| \leq s$.*

*Proof.* Let $\phi$ be a model of $H' \setminus E(H'_*)$ in $G_1^{\mathcal{C}}$, as in Definition 4.7. Since $G_1^{\mathcal{C}}$ is a subgraph of $G'$, $\phi$ can be considered as a model of $H' \setminus E(H'_*)$ in $G'$. We try to extend $\phi$ to a model of $H'$ in $G'$ by assigning values to $\phi(e)$ for every $e \in E(H'_*)$. In order to do this, let us make every vertex of $U \cup S^{\mathcal{C}}$ a root $G_2^{\mathcal{C}}$, and let $H''_*$ be obtained from $H'_*$ by making every vertex $v$ a root with the same root number as $\psi(v)$. We try to find a $w^{\mathcal{C}}$-bounded model $\psi$ of $H'_*$ in $G_2^{\mathcal{C}}$. Note that Definition 4.7 ensures that such a $\psi$ respects the degree condition: for every $v \in V(H'_*) \cap V(H)$, we have $\psi(v) \in U \cup S^{\mathcal{C}}$ and hence $w^{\mathcal{C}}(\psi(v)) = \delta$, while the degree of every $v \in V(H'_*) \setminus V(H)$ is at most 1 in $H'_*$. We use Lemma 4.9 to find either a small separation $(G'_1, G'_2)$, or a model $\psi$ of $H''_*$ in $G_2^{\mathcal{C}}$ with $\psi(v) = \phi(v)$ for every $v \in V(H)$. If Lemma 4.9 gives us a separation, then we are done. Otherwise, let us set $\phi(e) = \psi(e)$ for every $e \in E(H'_*)$. The paths $\phi(e)$ for $e \in E(H'_*)$ are pairwise internally disjoint: this follows from the fact that if $e \in E(H'_*)$, then the internal vertices of $\phi(e) = \psi(e)$ are in $V(G_2^{\mathcal{C}})$, while for every $e \in E(H') \setminus E(H'_*)$, the internal vertices of $\phi(e)$ are not in $V(G_2^{\mathcal{C}})$ (by Definition 4.7(3)). Thus $\phi$ is indeed a model of $H'$. □

Having established the correspondence between topological minors and tuples of clusters realizing a scheme, we concentrate on finding such a tuple. We observe that only the equivalence types of the clusters matter:

**Proposition 4.11.** *Let $H$ be a rooted graph and $(H', H'_*)$ be a scheme of $H$. Let $(C_{i_1}, \ldots, C_{i_r})$ and $(C_{i'_1}, \ldots, C_{i'_r})$ be two r-tuple of clusters such that $(C_{i_1}, \ldots, C_{i_r})$ realizes $(H', H'_*)$ and for every $1 \leq j \leq r$, clusters $C_{i_j}$ and $C_{i'_j}$ are equivalent. Then $(C_{i'_1}, \ldots, C_{i'_r})$ also realizes $(H', H'_*)$.*

The following lemma is standard: it shows that finding small fixed-size "colorful" independent sets in bounded-degree graphs can be done in linear time.

**Lemma 4.12.** *Let $W$ be a graph with maximum degree $d$ where the vertices are labeled with $k$ different labels. We can find in time $f(d,k) \cdot (|V(W)| + |E(W)|)$ an independent set of size $k$ where every vertex has a different label (or correctly state that there is no such set).*

**Lemma 4.13.** *Given a scheme $(H', H'_*)$ with $|V(H')| \leq 4\delta + 2\delta^2$, in time $f(\delta, L)|V(G)|$ (for some function $f(\delta)$) we can find a tuple $\mathcal{C} = (C_{i_1}, \ldots, C_{i_r})$ of clusters with $r \leq 2\delta$ that realizes $(H', H'_*)$ (if such a tuple exists).*

*Proof.* Let us enumerate all clusters and sort them into equivalence classes (where equivalence is understood according to Definition 4.4). Let $t$ be the number of equivalence classes and let us assign an integer $\tau(C_i) \in [t]$ to each cluster $C_i$ based on which class it belongs to. For every subset $T \subseteq [t]$ of size at most $2\delta$, we test whether there is a tuple $(C_{i_1}, \ldots, C_{i_{|T|}})$ of pairwise independent clusters with $\{\tau(C_{i_1}), \ldots, \tau(C_{i_{|T|}})\} = T$. In order to do this, we build a graph $W_T$ by introducing a vertex with label $\tau(C_i)$ corresponding to every cluster $C_i$ with $\tau(C_i) \in T$. Two vertices of $W_T$ are adjacent if the corresponding clusters are *not* independent. We claim that the maximum degree of $W_T$ can be bounded by a function of $\delta$ and $L$. To see this, recall that the maximum degree of $G \setminus U$ is at most $L$ and that the maximum distance in $G \setminus U$ between two vertices of a cluster $C_i$ is $O(\delta L)$ (as $C_i$ induces a connected subgraph of $F$). Thus if $C_i$ and $C_j$ are not independent, then $C_j$ is fully contained in the $O(\delta L)$-neighborhood of every vertex of $C_i$; the number of such sets can be bounded by a function of $\delta$ and $L$. This means that if we use Lemma 4.12 to find a colorful independent set in $W_T$, then the running time is linear in the number of clusters (for fixed $\delta$ and $L$). If Lemma 4.12 returns an independent set, then we test if the corresponding pairwise independent tuple $\mathcal{C} = (C_{i_1}, \ldots, C_{i_{|T|}})$ of clusters realizes $(H', H'_*)$ (as the size of $G_1^{\mathcal{C}}$ is bounded by a function of $\delta$ and $L$, this can be done by brute force). If after trying every $T \subseteq [t]$ of size at most $2\delta$, no tuple realizing $(H', H'_*)$ is found, then by Proposition 4.11 we know that there is no tuple realizing $(H', H'_*)$. $\square$

In Case 2 ($|U| < h$), our algorithm for FINDIRRELEVANTORSEPARATION determines the $\delta$-folio of $G' = G + X$ the following way. For every candidate $H$ in the $\delta$-folio, we enumerate every scheme $(H', H'_*)$ of $H$ with $|V(H)| \leq 4\delta + 2\delta^2$ (the number of such schemes is clearly bounded by a function of $\delta$). For each scheme, we use Lemma 4.13 to check if there is a tuple of clusters that realizes this scheme. If there is such a tuple, then by Lemma 4.10, we can obtain a model of $H$ in $G'$ or a small separation; if there is no such tuple, then the (contrapositive) of Lemma 4.8 shows that $H'$ is not a topological minor of $G'$. It is easy to verify that for fixed $\delta$ and $L$, the running time is $O(|V(G)|^2)$.

## 5 No clique case

It remains to prove Theorem 3.3. Let us recall the statement.

> There is an algorithm satisfying the specification of FINDIRRELEVANTORCLIQUE that runs in $f_1(\delta, |R(G)|, k) \cdot |V(G)|$ steps, for some computable function $f_1$.

Throughout the proof of Theorem 3.3, we will have to analyze instances of topological minors. To do so, we will typically think of the topological minor as given as a set of internally disjoint paths. To make this explicit, we define a *path system* to be a set $\mathcal{P} = \{P_1, \ldots, P_k\}$ of internally disjoint paths. We allow that an element $P_i$ of a path system is trivial, however in this case, we require that $\not\exists j$ such that $V(P_i) \subseteq V(P_j)$. Thus every trivial path $P_i$ of $\mathcal{P}$ forms a 1-vertex component of the graph $\bigcup_{i=1}^{k} P_i$. A special type of path system which we will frequently consider is a *linkage* where the elements of the system are pairwise vertex disjoint. We will use $V(\mathcal{P})$ and $E(\mathcal{P})$ to refer to the vertex and edge sets of the graph $\bigcup_{i=1}^{k} P_i$. Finally, we say that a path system $\mathcal{Q}$ is *equivalent* to $\mathcal{P}$ if they have the same order and for every element $P \in \mathcal{P}$ there exists an element $Q \in \mathcal{Q}$ such that $P$ and $Q$ have the same endpoints.

This section is organized as follows. In the next subsection, we give a key result, the so-called "weak structure theorem", which also plays an important role in the the graph minor algorithm by Robertson and Seymour [11]. We will need a stronger version of the theorem with an additional property ensuring that for any "piece" of a topological minor in the structure, one can find many disjoint copies of this piece. The exact statement of this stronger version will require additional notation, which we present in the second subsection before stating the theorem. In Subsection 5.3, we state the main theorem, Theorem 5.9, of this section, and present the proof of Theorem 3.3 assuming Theorem 5.9. The proof of Theorem 5.9 will occupy the rest of the section. We introduce the Unique linkage theorem in Subsection 5.4. In Subsection 5.5, we give several technical lemmas on path systems in graphs almost embedded in the disc or in the cylinder. Finally, in Subsection 5.6, we give the proof of Theorem 5.9.

### 5.1 The weak structure theorem

For our proof, we will need to consider what Robertson and Seymour dubbed societies. A *society* is a pair $(G, \Omega)$ where $G$ is a graph and $\Omega$ is a cyclic ordering of a subset of the vertices of $G$. In a slight abuse of notation, we will use $\Omega$ to refer both to the set of vertices as well as the cyclic ordering.

We will often restrict our attention to societies which can be nearly embedded in the plane.

**Definition 5.1.** A society $(G, \Omega)$ *embeds in the disc up to 3-separations* if there exist pairwise edge disjoint subgraphs $G_0, G_1, \ldots, G_m$ for some non-negative integer $m$ which satisfy the following.

  i. $G = \bigcup_0^m G_i$ and $\Omega \subseteq V(G_0)$.
  ii. $|V(G_0)| \cap V(G_i)| \leq 3$ for all $1 \leq i \leq m$ and $V(G_i) \cap V(G_j) \subseteq V(G_0)$ for all $1 \leq i < j \leq m$.

*iii*. For all $1 \leq i \leq m$ and distinct vertices $x, y \in V(G_i) \cap V(G_0)$, there exists a path $P$ in $G_i$ from $x$ to $y$ intersecting $V(G_i) \cap V(G_0)$ exactly in the vertices $x$ and $y$.

  *iv*. If we let $D$ be the closed unit disc, then for all $1 \leq i \leq m$, there exist pairwise disjoint open discs $\Delta_i \subseteq D$ and an embedding $\sigma : G \hookrightarrow D - (\bigcup_1^m \Delta_i)$ such that

  a. the vertices of $G_0$ embedded on the boundary of $D$ are exactly the vertices of $\Omega$ in the cyclic order indicated by $\Omega$, and
  b. for every $1 \leq i \leq m$, the vertices of $G_0$ embedded on the boundary of $\Delta_i$ are exactly the vertices of $V(G_i) \cap V(G_0)$.

We define the graph $G_0'$, called the *closure of $G_0$*, to be the graph obtained from $G_0$ by adding an edge between every non-adjacent pair of vertices $u$ and $v$ such that there exists an index $i$ with $u, v \in V(G_0) \cap V(G_i)$.

We will often consider a society $(G, \Omega)$ embedded in the disc up to 3-separations with a minimal number of edges with respect to some desirable property. The next observation lays out when $G - e$ embeds up to 3-separations as well. The proof follows immediately from the definition and we omit it here.

*Remark* 5.2. Let $(G, \Omega)$ be a society which embeds in the disc up to 3-separations, and let $(\{G_0, G_1, \ldots, G_m\}, \sigma, \{\Delta_1, \ldots, \Delta_m\})$ be a fixed embedding up to 3-separations of $(G, \Omega)$. If $e \in E(G_0)$, then $(\{G_0 - e, G_1, \ldots, G_m\}, \sigma', \{\Delta_1, \ldots, \Delta_m\})$ is an embedding up to 3-separations of $(G - e, \Omega)$ where $\sigma'$ is the restriction of $\sigma$ to the graph $G_0 - e$. If $e \in E(G_i)$ for $i \in \{1, \ldots, m\}$, then $(\{G_0, G_1, \ldots, G_i - e, \ldots, G_m\}, \sigma, \{\Delta_1, \ldots, \Delta_m\})$ is an embedding up to 3-separations if and only if $e$ is not a cut edge separating vertices of $V(G_i) \cap V(G_0)$ in $G_i$. In general, for all $e \in E(G)$, the society $(G - e, \Omega)$ embeds in the disc up to 3-separations, although it might be necessary to modify a given embedding of $(G, \Omega)$ to obtain one for $(G - e, \Omega)$.

We will need some somewhat technical notation for describing sets of cycles in societies embedded up to 3-separations. Let $(G, \Omega)$ be a society and let $(\{G_0, G_1, \ldots, G_m\}, \sigma, \{\Delta_1, \ldots, \Delta_m\})$ be an embedding of the society in the disc up to 3-separations. A cycle $C$ in $G$ is *grounded* if $V(C) \cap V(G_0) \geq 3$. If we consider a grounded cycle $C$, then the subgraph $C \cap G_0$ is either a cycle where every edge of $C$ is contained in $G_0$ or it is a union of (possibly trivial) disjoint paths. In the case where it is a union of disjoint paths, we can label the components $P_0, \ldots, P_{l-1}$ for some positive integer $l$ and label the endpoints of $P_i$ as $x_i$ and $y_i$ for $1 \leq i \leq l$ such that the vertices $x_0, y_0, x_1, y_1, \ldots, x_{l-1}, y_{l-1}$ occur on $C$ in that order when traversing the cycle. In the case that $P_i$ is a trivial path, we let $x_i = y_i$. For every index $i$, $0 \leq i \leq l-1$, there exists an index $j$ and a path, call it $Q_i$, such that $x_i$ and $y_i$ are contained in $V(G_j) \cap V(G_0)$ and the path $Q_i$ is a subpath of $G_j$ linking $y_i$ and $x_{i+1}$ with the subscript notation taken modulo $l$. While it is possible that two paths $Q_i$ and $Q_{i'}$ will be contained in the same subgraph $G_j$ of the decomposition, they will be internally disjoint, possibly intersecting only in one vertex of $V(G_j) \cap V(G_0)$. We define the cycle $\bar{C}$ in the closure of $G_0$ obtained by restricting $C$ to the vertices of $G_0$ and replacing each missing subpath $Q_i$ contained in $C - E(G_0)$ by an edge of $E(G_0') \setminus E(G_0)$, where $G_0'$ again is the closure of $G_0$. Call $\bar{C}$ the *projection of $C$ to $G_0'$*. Note that the requirement that $|V(C) \cap V(G_0)| \geq 3$ ensures that $\bar{C}$ is a well defined cycle of $G_0'$.

For the grounded cycle $C$, the projection $\bar{C}$ of $C$ to $G_0'$ defines a closed subdisc $\Delta^c$ of the disc. Let $G_0^c$ be the induced subgraph of $G_0$ with vertices in $\Delta^c$. We let $m'$ be a positive integer and $G_1^c, \ldots, G_{m'}^c$ be the (re-labeled) set of $\{G_i : V(G_i) \cap V(G_0) \subseteq (V(G_0^c))\}$. Let $G^c$ be the subgraph of $G$ given by $G^c = G_0^c \cup \bigcup_{i=1}^{m'} G_i^c$. Let $\Omega^c$ be the natural cyclic order of $V(\bar{C})$. We refer to the society $(G^c, \Omega^c)$ as the *embedding-induced society* of the grounded cycle $C$. Observe that the embedding up to 3-separations of $(G, \Omega)$ immediately yields an embedding up to 3-separations of $(G^c, \Omega^c)$. Note, it is possible that the cycle $C$ is not contained in $G^c$, specifically when there exists a $G_j$ with $|V(G_j) \cap V(G_0)| = 3$ containing a subpath of $C$ such that exactly two vertices of $G_j$ are contained in $V(G^c)$ and one vertex is "outside" the disc bounded by $\bar{C}$. In this case, if we let $\Omega_2^c$ be the natural cyclic order of the vertices $V(C)$ given by $C$, then $(G^c \cup C, \Omega_2^c)$ has an embedding up to 3-separations as well.

We will be specifically interested in embeddings up to 3-separations which contain a large grid-like graph. Towards this end, we now consider *walls*.

For positive odd integers $r$, define a graph $H_r$ (called the *basic r-wall*) as follows. Let $P_0,\ldots,P_r$ be $r$ vertex disjoint paths of length $2r+1$, say $P_i = v_0^i \ldots v_{2r+1}^i$. Let $V(H_r) = \bigcup_{i=1}^r V(P_i) \setminus \{v_0^0, v_{2r+1}^r\}$, and let

$$E(H_r) = \left(\bigcup_{i=1}^r E(P_i) \setminus \{v_0^0 v_1^0, v_{2r}^r v_{2r+1}^r\}\right) \cup \left\{v_j^i v_j^{i+1} : i \text{ odd}, j \text{ even}; 1 \leq i < r; 0 \leq j \leq 2r+1\right\}$$

$$\cup \left\{v_j^i v_j^{i+1} : i \text{ even}, j \text{ odd}; 0 \leq i < r; 1 \leq j \leq 2r+1\right\}.$$

Note that the restriction of $r$ to be even ensures that the indices are more nicely behaved. The 6-cycles in $H_r$ are its *bricks*. In the natural plane embedding of $H_r$, these bound its 'finite' faces. The outer cycle of the unique maximal 2-connected subgraph of $H_r$ is the *boundary cycle* of $H_r$. The paths $P_0,\ldots,P_r$ are called the *horizontal paths* of the wall. Let the path $Q_i$ for $0 \leq i \leq r$ be the path of $H_r$ induced by the set of vertices $\{v_j^{2i}, v_j^{2i+1} : 1 \leq j \leq r-1\} \cup \{v_0^{2i}, v_r^{2i+1}\}$. The $Q_i$ will be called the *vertical paths* of the wall.

The cycle $C$ of $H_r$ is a *rectangle* if $C$ is a subgraph of the union of exactly two horizontal and two vertical paths of $H_r$. Note that the bricks as well as the boundary cycle of $H_r$ are rectangles. The *diameter* of a rectangle contained in the union of $P_i, P_{i'}, Q_j, Q_{j'}$ is the $\max\{|i-i'|, |j-j'|\}$. Thus the rectangles of diameter 1 are exactly the bricks of $H_r$.

Any subdivision $H$ of $H_r$ will be called an *r–wall* or a *wall of size $r$*. The bricks and boundary cycle of $H$ are its subgraphs that form subdivisions of the bricks and boundary cycle, respectively. Recall that to dissolve a vertex of degree 1 or 2 in a graph, we simply contract an incident edge. Given such a wall $H$ of size $r$, let $X$ be a subset of vertices of $H$ containing every vertex of degree 3 in $H$ such that dissolving every vertex of degree two in $V(H) \setminus X$ results in the graph $H_r$. Call such a set $X$ of vertices *pegs* of the wall. We can label the set of pegs $v_i^j$, $0 \leq i \leq 2r+1$ $0 \leq j \leq r$ according to their position in the wall after suppressing the vertices of degree two in $V(H) \setminus X$. Call such a labeling the *canonical labeling of the pegs*. Given a set of pegs of a wall $H$, the horizontal and vertical paths of a wall $H$ are the subdivided paths of $H_r$ corresponding to the horizontal and vertical paths of $H_r$. We let the rectangles be the cycles of $H$ forming the subdivisions of the rectangles of $H_r$.

**Definition 5.3.** Let $r$ and $t$ be positive integers. A *nearly flat r-wall decomposition with apex bound $t$* of a graph $J$ is given by a 5-tuple $(A, G, H, W, X)$ such that $A \subseteq V(J)$, $G$ and $H$ are subgraphs of $J$, $W$ is an $r$-wall in $J$ and $X$ is a set of pegs of $W$ which satisfy the following. Let $C$ be the boundary cycle of $W$.

i. $J - A = G \cup H$ and $W$ is a subgraph of $G$.
ii. $V(H) \cap V(G) \subseteq V(C)$, and if we let $\Omega$ be the natural cyclic order of $V(H) \cap V(G)$ given by the cycle $C$, the society $(G, \Omega)$ has an embedding up to 3-separations $\{\{G_0, G_1, \ldots, G_m\}, \sigma, \{\Delta_1, \ldots, \Delta_m\}\}$.
iii. The set $X$ of pegs is contained in $V(G_0)$.
iv. $|A| \leq t$ and for every $x \in A$ and every brick $B$ of $W$, $x$ has a neighbor $y$ in the embedding induced society of the brick $B$ and there exists a path from $y$ to $B$ in the embedding induced society.

The goal of this subsection is to present a theorem saying when we can find a large nearly flat wall decomposition. Essentially, we will see that we can always find such a decomposition or the graph must have *bounded tree width* or a large *clique minor*. We remind the reader that a graph $G$ contains $K_t$ as a minor if there exist pairwise disjoint subsets $X_1, \ldots, X_t$ of vertices such that $G[X_i]$ is connected for all $1 \leq i \leq t$ and for all $1 \leq i < j \leq t$, there exists an edge with one endpoint in $X_i$ and one endpoint in $X_j$. These sets are referred to as the *branch sets* of the minor.

Finally, we will refer to the *treewidth* of a graph $G$, denoted $tw(G)$. However, we will not need the technical definition here and so omit it.

We are now ready to give the weak structure theorem.

**Theorem 5.4 (Weak Structure Theorem, [11], Theorem (9.4)).** *For all $t \geq 1$, $r$ even, there exists a value $w = w(t, r)$ such that the following holds. Let $J$ be a graph on $n$ vertices of treewidth at least $w$. There exists an $O(|V(G)|)$ time algorithm that outputs either sets of vertices $\{X_1, X_2, \ldots, X_t\}$ forming a $K_t$ minor or outputs $(A, G, H, W, X)$ forming a nearly flat r-wall decomposition with apex bound $t^2$.*

Note that Robertson and Seymour [11] gives an $O(V|(G)|^2)$ time algorithm to output either the nearly-flat $r$-wall decomposition or a $K_t$ minor. The time complexity is improved in [8] to $O(|V(G)|)$.

## 5.2 Strengthening Theorem 5.4

In this subsection, we present a strengthening of the weak structure theorem which will allow us to find an irrelevant vertex.

We will first need some notation describing sets of cycles in societies embedded up to 3-separations.

**Definition 5.5.** Let $(G, \Omega)$ be a society and let $(\{G_0, G_1, \ldots, G_m\}, \sigma, \{\Delta_1, \ldots, \Delta_m\})$ be an embedding of the society in the disc up to 3-separations. An *s-nest* is a set $\mathcal{C} = \{C_1, C_2, \ldots, C_s\}$ of disjoint cycles which satisfy the following.

 i. For all $1 \le i \le s$, $C_i$ is grounded.
 ii. For all $1 \le i \le s$, let $\bar{C}_i$ be the projection of $C_i$ to $G'_0$. Then for all $j > i$, the vertices $V(\bar{C}_j)$ are contained in the subdisc bounded by $\bar{C}_i$. [1]

As with path systems, we will use $E(\mathcal{C})$ to refer to the edge set $\bigcup_{C \in \mathcal{C}} E(C)$ and $V(\mathcal{C})$ for the vertex set $\bigcup_{C \in \mathcal{C}} V(C)$.

**Definition 5.6.** Let $\mathcal{C} = \{C_1, \ldots, C_s\}$ be a set of $s$ disjoint cycles in a graph $G$. We say the path system $\mathcal{P} = \{P_1, \ldots, P_k\}$ is *perpendicular to* $\mathcal{C}$ if the following conditions hold for all $1 \le i \le k$.

 i. For all $i$ and $j$, if $V(P_i) \cap V(C_j) \ne \emptyset$, then for all $j'$, $1 \le j' \le s$, $V(P_i) \cap V(C_{j'}) \ne \emptyset$.
 ii. There does not exist an element $P_i \in \mathcal{P}$ containing vertices $x, y, z$ such that when traversing $P_i$ from one end to the other we encounter $x, y, z$ in that order and distinct indices $j, j'$ such that $x, z \in V(C_j)$ and $y \in V(C_{j'})$.
 iii. For all indices $i$, $1 \le i \le k$ and for all $j$, $1 \le j \le s$, $P_i \cap C_j$ is a (non-empty) subpath of $C_j$.

We say that $\mathcal{P}$ is *nearly perpendicular* if it satisfies *i* and *ii*.

We will need to define a canonical set of concentric cycles for each rectangle. Let $(A, G, H, W, X)$ be a nearly flat wall decomposition, and let $C$ be a rectangle of $W$, and let $d$ be a positive integer. The *d-target centered at $C$* is a $d$-nest $\{C_1, C_2, \ldots, C_d\}$ satisfying the following

 i. For all $1 \le i \le d$, $C_i$ is a rectangle of $W$.
 ii. For all $1 \le i \le d$, the embedding induced society of $C_i$ contains both $C$ and $C_j$ for all $j > i$.
 iii. With respect to *i* and *ii*, $C_1, \ldots, C_d$ are chosen to minimize the embedding induced society of $C_d$.

Thus, a $d$-target centered at $C$ can be thought of as the next $d$ rectangles surrounding $C$ in the wall decomposition. Note that the $d$-target is in fact uniquely determined by $C$.

We are now ready to present the additional property which we will add to Theorem 5.4. We define what we will call a pattern in a nearly flat wall decomposition. A pattern can be thought of as a piece of a topological minor in a graph admitting a nearly flat wall decomposition such that this piece has the additional property that it intersects nicely with a given nest contained in the wall.

**Definition 5.7.** Let $r$, $l$, and $k$ be positive integers. Let $J$ be a graph, and let $(A, G, H, W, X)$ be a nearly flat $r$-wall decomposition of $J$. Let $C$ be a rectangle. A *pattern centered at $C$* of order $k$ and depth $l$ consists of a path system $\mathcal{P}$ of order $k$ satisfying the following properties. Let $\{C_1, \ldots, C_l\}$ be the $l$-target centered at $C$. Let $J_C$ be the subgraph of $J$ given by the union of the embedding induced society of $C$ and all the edges with one endpoint in the embedding induced society and one endpoint in $A$. Similarly define $J_{C_1}$. Thus $J_{C_1}$ contains all the $l$-target $\{C_1, \ldots, C_l\}$.

 i. The path system $\mathcal{P}$ is contained in the subgraph $J_{C_1}$.
 ii. For all $P \in \mathcal{P}$, $P$ has no internal vertex contained in $A$.
 iii. For all $P \in \mathcal{P}$, if $P \cap (V(J_{C_1}) \setminus V(J_C)) \ne \emptyset$ then $P$ has exactly one endpoint in $V(J_C)$, the other endpoint in $V(C_1)$, and $P$ is perpendicular to $\{C_1, C_2, \ldots, C_l\}$. [2]

---
[1] Thus the "inner" cycle is $C_s$ and the "outer" cycle is $C_1$.
[2] Note we have to modify the definition of perpendicular to allow paths that don't have both endpoints contained in the nest

Note, specifically we require that the $l$-target centered at $C$ be defined.

Let $l$ and $k$ be positive integers and let $\mathcal{P}$ and $\mathcal{P}'$ be patterns of order $k$ and depth $l$ centered at the rectangles $C$ and $C'$, respectively. Let the set of endpoints of elements of $\mathcal{P}$ and $\mathcal{P}'$ be $S$ and $S'$, respectively. We say that $\mathcal{P}$ is *homeomorphic* to $\mathcal{P}'$ if there exists bijections $\pi_1 : S \to S'$ and $\pi_2 : \mathcal{P} \to \mathcal{P}'$

  i. For all $x \in (S \cup S') \cap A$, $\pi_1(x) = \pi_1^{-1}(x) = x$.
  ii. For all $P \in \mathcal{P}$ with endpoints $x$ and $y$, the endpoints of $\pi_2(P)$ are $\pi_1(x)$ and $\pi_1(y)$.

Moreover, if we fix an orientation of the embedding of $G$ in the plane and label it clockwise, we have the following property.

  iii. If we let $x_1, \ldots, x_t$ be the vertices of $S \cap V(C_1)$ so that they occur in that clockwise order on $V(C_1)$, then $\pi_1(x_1), \ldots, \pi_1(x_t)$ occur on $C_1'$ in that clockwise order.

We are now give the following strengthening of the weak structure theorem.

**Theorem 5.8.** *For all positive integers $t, \delta, l, d \geq 1$ and $r$ even, there exists a value $w = w(t, \delta, l, d, r)$ such that the following holds. Let $J$ be a graph on $n$ vertices with $tw(J) \geq w$. There exists an $O(|V(G)|)$ time algorithm that outputs either the branch sets of a $K_t$ minor or outputs $(A, G, H, W, X)$ forming a nearly flat $r$-wall decomposition with apex bound $t^2$ with the following property.*

  v. *For all rectangles $C$ of diameter at most $d$ admitting an $l$-target and for every pattern $\mathcal{P}$ centered at $C$ of depth $l$ and order at most $\delta$ we have that for every brick $B$ admitting an $l$-target there exists a pattern $\mathcal{P}'$ centered at $B$ of depth $l$ which is homeomorphic to $\mathcal{P}$.*

Again, Robertson and Seymour [11] give an $O(V|(G)|^2)$ time algorithm to output the structure in Theorem 5.8; the complexity was improved to $O(|V(G)|)$ in [8]. The structure guaranteed by Theorem 5.8 is very similar to the one defined in [11], which is called a "homogeneously labeled wall". This wall has many "similar", disjoint subwalls, each of which can play an equivalent role with respect to the folio. The exact statement of Property v. in Theorem 5.8 is derived to be more friendly to maintaining topological subgraphs. An algorithm to construct the structure given in Theorem 5.8 can be easily obtained from Theorem 5.4. In fact, (10.1) in [11] is almost exactly the statement above, however we must modify the structure slightly in order to ensure that condition v. holds. Given the structure in Theorem 5.4 and the algorithm of (10.1) in [11], it is straightforward to obtain an $O(n)$ time algorithm to find the structure as in Theorem 5.8 (the statement follows from a typical Ramsey type argument as in the proof of (10.1) in [11]. We omit the proof here).

## 5.3 Proof of Theorem 3.3

In this subsection, we present the proof of Theorem 3.3. Our primary tool in doing so is the following which connects the structure given in Theorem 5.8 to the existence of irrelevant vertices. Recall from Section 3 that a set $X$ of vertices is *irrelevant* to the $\delta$-folio of $G$, if rooted graphs $G$ and $G \setminus X$ have the same $\delta$-folio. We say that a vertex $v$ is irrelevant if the set $\{v\}$ is irrelevant.

**Theorem 5.9.** *Let $\delta$ and $t$ be positive integers. There exist values $r = r(\delta, t)$, $d = d(\delta, t)$, and $l = l(\delta, t)$, which satisfy the following. Let $J$ be a rooted graph with roots $R(J)$. Assume that $J$ admits a nearly flat $r$-wall decomposition $(A, G, H, W, X)$ with apex bound $t$ satisfying properties i-v. Assume that $R(J) \cap V(G) = \emptyset$. Finally, let the pegs have the canonical labeling. Then the peg $v_r^{r/2}$ is irrelevant to the $\delta$-folio of $J$.*

Note that algorithmically, given the $r$-wall decomposition in the statement of Theorem 5.9, it is trivial to output the peg $v_r^{r/2}$ in constant time. The proof of Theorem 5.9 will occupy the remainder of this section; we delay the proof until the later subsections and continue with the proof of Theorem 3.3.

The second result which we will need is the following, showing the $\delta$-folio can be solved in polynomial time if the treewidth is bounded.

**Theorem 5.10 (See [1, 11]).** *For integers $w$ and $\delta$, there exists a $(w+\delta)^{O(w+\delta)}O(|V(G)|)$ time algorithm for computing the $\delta$-folio in graphs of treewidth $w$.*

We are now give the algorithm for Theorem 3.3, satisfying the specifications of FINDIRRELEVANTORCLIQUE defined in Section 3.

Let $J$ be our graph with roots $R(J)$ and constants $\delta$ and $k$ be given as input. Fix $r = r(\delta, k^2)$, $d = d(\delta, k^2)$, and $l = l(\delta, k^2)$ from Theorem 5.9. Let $r' = |R(J)|r$, and let $w = w(k, \delta, d, l, r')$ as in Theorem 5.8.

**Step 1. The small treewidth case.** As a first step, test whether or not $J$ has treewidth at least $w$. This can be done by the algorithm of Bodlaender [2]. If $J$ does have $tw(J) \leq w$, then apply Theorem 5.10 to obtain the $\delta$-folio. Otherwise, go to Step 2.

**Step 2. Apply the weak structure theorem.** Apply Theorem 5.8 to $J$ with $t = k$. The algorithm either outputs the branch sets of a $K_k$ minor, or we find the structure $(A, G, H, W, X)$ forming a flat $r'$-wall decomposition with apex bound $k^2$. Go to Step 3.

**Step 3. Find an irrelevant vertex.** We would like to apply Theorem 5.9 to output an irrelevant vertex. As a final technicality, it is possible that there are roots of $R(J)$ contained in the subgraph $G$ of the decomposition. However, by our choice of $r'$, there exists an $r$-subwall $W'$ of $W$ with boundary cycle $C'$ such that the embedding induced society $G'$ of $C'$ does not contain any vertices of $R(J)$. Let $H'$ and $X'$ be accordingly defined so that $(A, G', H', W', X')$ forms a flat $r$-wall decomposition with $R(J) \cap V(G') = \emptyset$. Note that we can find such a subwall $W'$ in linear time. Apply Theorem 5.9 to the decomposition $(A, G', H', W', X')$ to obtain an irrelevant vertex, which we then output.

Let us clarify the time complexity of this algorithm. Step 1 can be done in $O(|V(G)|)$ time because we apply Theorem 5.10 and the algorithm of Bodlaender [2], and both run in $O(|V(G)|)$ time. Step 2 can be done in $O(|V(G)|)$ time by Theorem 5.8. Step 3 can be done in $O(|V(G)|)$ time by Theorem 5.9. Thus the overall runtime is $O(|V(G)|)$, as desired.

Thus, all that remains is to prove Theorem 5.9. We do so in several steps. In the next two subsections, we give several auxiliary results before presenting the proof of Theorem 5.9 in subsection 5.6. Let us emphasize here that our proof of Theorem 5.9 does not depend on the full power of the graph minor structure theorem [12]. We can avoid the structure theorem, because there is now a shorter proof for the correctness of the graph minor algorithm in [9]. Utilizing some results in [9], we are able to avoid the much of the heavy machinery of the graph minor structure theory.

## 5.4 Unique Linkage Theorem

Our primary tool in the next subsection will be a powerful theorem of Robertson and Seymour known as the Unique linkage theorem [13].

**Theorem 5.11 ([13]).** *For all $k \geq 1$, there exists an integer $w(k)$ satisfying the following. Let $G$ be a graph and $\mathcal{P}$ a linkage of order $k$ contained in $G$ such that $V(G) = V(\mathcal{P})$. If $tw(G) \geq w(k)$, then there exists a vertex $v \in V(G)$ and a linkage $\mathcal{P}'$ equivalent to $\mathcal{P}$ with $V(\mathcal{P}') \subseteq V(G) - v$.*

To describe the existence of such a vertex $v$ and linkage $\mathcal{P}'$ as in Theorem 5.11, we will often say that the path system $\mathcal{P}$ *can be re-routed to avoid* some vertex $v$ of $G$. We will need to apply Theorem 5.11 in a slightly more general context. Towards that end, we give the following corollary.

**Corollary 5.12.** *For all $k \geq 1$, there exists an integer $w(k)$ satisfying the following. Let $G$ be a graph and $\mathcal{P}$ a path system of order $k$ contained in $G$. Let $Z = \{v \in V(G) : \deg(v) \geq 3\}$. Assume that $Z \subseteq V(\mathcal{P})$. If $tw(G) \geq w(k)$, then there exists a vertex $v \in Z$ and a path system $\mathcal{P}'$ equivalent to $\mathcal{P}$ with $V(\mathcal{P}') \subseteq V(G) - v$.*

*Proof.* Assume the claim is false, and let $G$ and a path system $\mathcal{P}$ of order $k$ be a counterexample. Let $S$ be the set of endpoints of elements of $\mathcal{P}$. Let $w(k)$ be the function given in Theorem 5.11. We assume the treewidth of $G$ is at least $w(k)$. Assume that from all such counterexamples, we pick a counterexample minimizing $2k - |S|$ and subject to this, we pick a counter example on a minimum number of edges.

First, observe that if there exists a vertex $v$ of degree one or two in $V(G) \setminus S$, then if we let $\bar{G}$ be the graph obtained by dissolving $v$ and $\bar{\mathcal{P}}$ the path system obtained by dissolving $v$, then by our choice of counterexample, there exists a path system equivalent to $\bar{\mathcal{P}}$ avoiding some vertex of degree three. This path system will correspond to a linkage in $G$ which is equivalent to $\mathcal{P}$ and avoids some vertex of degree 3 as well, a contradiction. Thus we see that $V(G) = V(\mathcal{P})$ and that every vertex of $V(G) \setminus S$ has degree at least 3.

Note that if $2k - |S| = 0$, then $\mathcal{P}$ is a linkage and the claim follows from Theorem 5.11. Thus, we see that $2k - |S| \geq 1$ and that there exist two elements $P_1$ and $P_2$ of $\mathcal{P}$ sharing a common endpoint $v$. Let $G'$ be the graph obtained by deleting the vertex $v$ and adding two vertices $v_1$ and $v_2$ each adjacent in $G'$ to every vertex of the neighborhood in $G$ of the vertex $v$, i.e. $N_{G'}(v_1) = N_{G'}(v_2) = N_G(v)$. Let $P_1'$ be the path in $G'$ obtained by adding the vertex $v_1$ to the subpath $P_1 - v$. For all $P \in \mathcal{P} - P_1$, if $v$ is an endpoint of $P$, let $P'$ be the path of $G'$ obtained by adding $v_2$ to the subpath $P - v$. If $v$ is not an endpoint of $P \in \mathcal{P} - P_1$, let $P' = P$, and let $\mathcal{P}' = \{P' : P \in \mathcal{P}\}$. By construction, the path system $\mathcal{P}$ can be rerouted in $G$ to avoid some vertex if and only if the path system $\mathcal{P}'$ can be rerouted in $G'$ to avoid some vertex as well.

The graph $G$ is a subgraph of $G'$, and so we have that $tw(G') \geq tw(G) \geq w(k)$. Moreover, if we let $S'$ be the set of endpoints of elements of $\mathcal{P}'$, we see that $|S'| = |S| + 1$. Thus, $2k - |S'| < 2k - |S|$, contradicting our choice of counterexample. This proves the claim. □

## 5.5 Routing for discs and cylinders

In this subsection, we will give several technical lemmas concerning almost planar graphs embedded in the disc and cylinder. These lemmas look at how path systems intersect large societies embedded in the disc up to 3-separations. Specifically, we will see how large nests in the embedded societies allow us to reroute the path systems to achieve certain desirable properties.

First, we make a simple observation on nests in embedded societies. For any $s$-nest $\mathcal{C} = \{C_1, \ldots, C_s\}$ in a society $(G, \Omega)$ with an embedding in the disc up to 3-separations $(\{G_0, G_1, \ldots, G_m\}, \sigma, \{\Delta_1, \ldots, \Delta_m\})$, we have the following property. Let $G^i = G^{c_i}$ and $\Omega^i = \Omega^{c_i}$ with $(G^i, \Omega^i)$ equal to the embedding induced society of $C_i$. While it is certainly possible that the cycle $C_i$ will not be contained in $G^i$, we do have that $C_j$ will be a subgraph of $G^i$ for all $i < j \leq s$. It is an easy observation that $\mathcal{C}' = \{C_{i+1}, \ldots, C_s\}$ form an $(s-i)$-nest in the society $(G^i, \Omega^i)$ with the natural induced embedding up to 3-separations.

We will need the concept of a bramble which certifies when a graph has large treewidth. Given two subgraphs $H_1$ and $H_2$ of a graph $G$, we say that $H_1$ and $H_2$ *touch* if either there exists an edge with one end in $V(H_1)$ and the other end in $V(H_2)$, or alternatively, the subgraphs have a vertex in common. A *bramble* is a set of pairwise touching, connected subgraphs. A subset $X \subseteq V(G)$ *covers* a bramble if every element contains a vertex of $X$. The *order* of a bramble $\mathcal{B}$, denoted $ord(\mathcal{B})$, is the minimum size of a cover of the bramble. The next theorem of Seymour and Thomas [15] shows the relationship between the size of a bramble and the treewidth of a graph.

**Theorem 5.13 ([15]).** *Let $G$ be a graph. Then*

$$\max_{\mathcal{B} \text{ is a bramble}} ord(\mathcal{B}) = tw(G) + 1.$$

We will be considering a similar set-up in the following lemmas. We formalize it in the following common hypothesis.

> **Hypothesis 1.** Let $G$ and $H$ be graphs, and let $s$ be a positive integer. Let $\Omega$ be a cyclic ordering of the vertices of $V(H) \cap V(G)$. Let $W$ be the graph $G \cup H$. Let $(\{G_0, G_1, \ldots, G_m\}, \sigma, \{\Delta_1, \ldots, \Delta_m\})$ be an embedding of $(G, \Omega)$ in the disc $\Delta$ up to 3-separations. Let $\mathcal{C} = \{C_1, \ldots, C_s\}$ be an $s$-nest in $(G, \Omega)$. For $1 \leq i \leq s$, let $(W_i, \Omega_i)$ be the embedding induced society of $C_i$.

We will need one more definition before proceeding. Let $P$ and $Q$ be paths such that $Q$ has both endpoints on $P$. Let the ends of $P$ be $x_P, y_P$ and the ends of $Q$ be $x_Q, y_Q$ and assume $x_P, x_Q, y_Q, y_P$ occur on $P$ in that order. Then the path obtained by *rerouting P through Q* is the path $x_P P x_Q Q y_Q P y_P$.

The following lemma essentially shows that given an almost embedded planar graphs embedded in the disc with a large number of nested cycles and a linkage with all its endpoints contained outside the nested cycles, then we can re-arrange the linkage so that no path hits a deeply nested cycle.

**Lemma 5.14.** *Let $t,k$ be positive integers, and let $w(k)$ be the value given by Corollary 5.12. Assume Hypothesis 1. Let $\mathcal{P}$ be a path system of order $k$ in $W$ such that for every $P \in \mathcal{P}$, the endpoints of $P$ are contained in $V(H)$. If $s \geq t + 2w(k) + 2$, then there exists a path system $\mathcal{P}'$ equivalent to $\mathcal{P}$ such that $\mathcal{P}'$ is disjoint from $W_{s-t}$.*

*Proof.* We assume the lemma is false, and pick a counter-example containing a minimal number of edges. Thus, we may assume that there does not exist any edge of $G$ which can be deleted without changing the embedding up to 3-separations of the graph.[3] It immediately follows from Remark 5.2 that $E(G_0) \subseteq \bigcup_{i=1}^{s} E(C_i) \cup \bigcup_{P \in \mathcal{P}} E(P)$. Our first claim below will look at when some graph $G_i$ can contain edges as well which are not contained in $E(C_i) \cup \bigcup_{P \in \mathcal{P}} E(P)$. We now add the assumption that, with respect to containing a minimal number of edges, we pick an embedding with $m$ minimized.

*Claim* 1. The value $m$ is 0, i.e. $G$ is planar and $G_0 = G$.

*Proof.* Assume $m \geq 1$ and consider $G_m$. As $|V(G_0) \cap V(G_m)| \leq 3$, we see that at most one cycle of $\mathcal{C}$ and at most one element of $\mathcal{P}$ contains an edge of $E(G_m)$. Let $T$ be a spanning tree of $G_m$ containing $(\bigcup_{i=1}^{s} E(C_i)) \cap E(G_m)$. If $E(\mathcal{P}) \cap E(G_m) \neq \emptyset$, we let $\mathcal{P}'$ be a linkage equivalent to $\mathcal{P}$ such that $E(G_m) \cap E(z\mathcal{P}') \subseteq E(T)$. Otherwise, we let $\mathcal{P}' = \mathcal{P}$. If there exists an edge $e$ of $E(G_m) \setminus E(T)$, then it follows that $G - e$ with the embedding up to 3-separations $(\{G_0, G_1, \ldots, G_m - e\}, \sigma, \{\Delta_1, \ldots, \Delta_m\})$ violates our choice of counterexample to contain a minimal number of edges.

We conclude that $G_m = T$. However, in this case, we can embed $G_m$ in the disc $\Delta_m$ with the vertices of $V(G_m) \cap V(G_0)$ on the border. Thus, $G_0 \cup G_m$ embeds in $\Delta - (\bigcup_{i=1}^{m-1} \Delta_i)$ where $\Delta$ is the disc. As a technicality, if $C_i$ intersects $G_m$ in at least one edge, then in order to ensure that the subgraph $W_i$ remains unchanged, we need that $G_m$ embeds into $\Delta_m$ with the subpath $C_i \cap G_m$ on the boundary of the disc $\Delta_m$. Given that $G_m$ is a tree, this is possible. We conclude that the original embedding up to 3-separations violates our choice to minimize $m$, and consequently $m = 0$ proving the claim. $\diamond$

There are two important consequences of Claim 1. First, we see that $E(W) = \bigcup_{i=1}^{s} E(C_i) \cup \bigcup_{P \in \mathcal{P}} E(P)$. Secondly, it now follows that there does not exist an edge $e \in E(W) \setminus E(\mathcal{C})$ and a path system $\mathcal{P}'$ equivalent to $\mathcal{P}$ such that $e \notin E(\mathcal{P}')$, lest we again contradict minimality. We now show that the treewidth of $W$ is bounded by Corollary 5.12.

*Claim* 2. $tw(W) < w(k)$.

*Proof.* If the $tw(W) \geq w(k)$, then there exists a path system $\mathcal{P}'$ equivalent to $\mathcal{P}$ in $W$ and a vertex $v \in V(W) \setminus V(\mathcal{P}')$ by Corollary 5.12. Moreover, the vertex $v$ has degree 3, and so consequently there exists an edge $e$ of $W$ incident $v$ which is contained in $E(\mathcal{P}) \setminus E(\mathcal{C})$, a contradiction. $\diamond$

We define a *dive* to be a subpath $R$ contained in $\mathcal{P}'$ such that $R$ is an $\Omega$-path contained in $G$. Let $t' = \max_{1 \leq i \leq s} V(R) \cap V(C_i) \neq \emptyset$. We refer to $t'$ as the *depth* of the dive $R$.

*Claim* 3. For all $l \geq 2$, if there exists a dive of depth $l$, then there exists a dive of depth $l - 1$.

*Proof.* Consider a dive $R$ of depth $l$. The path $R$ in the disc $\Delta$ has both endpoints on the boundary of $\Delta$. Thus, it defines two closed sub discs of $\Delta$ intersecting in $R$. We fix $\Delta_R$ to be the sub-disc of $\Delta$ which does not intersect $C_{l+1}$ (when $l = s$, we fix $\Delta_R$ arbitrarily). We now fix $R$ to be a dive of depth $l$ minimizing $\Delta_R$ by inclusion. As $C_{l-1}$ intersects $\Delta_R$, we see that there exists a subpath $Q$ contained in $C_{l-1}$ with both endpoints in $R$ and no internal vertex in $R$. Assume that $P$ is the element of $\mathcal{P}$ containing $R$. Observe that there exists at least one edge of $P$, call it $e$, which is contained in the subpath of $R$ with both endpoints equal to the endpoints of $Q$ but is not contained in $E(\mathcal{C})$.

---

[3]The reason we would like to maintain the embedding is that the subgraphs $W_i$ for $1 \leq i \leq s$ are dependent on both the *s*-nest as well as the embedding. Further analysis could show that the embedding can be chosen so that the subgraph $W_{s-t}$ does not change; instead we limit ourselves to deleting edges which do not alter the embedding.

Assume as a case that $Q$ is internally disjoint from $\mathcal{P}$. Let $P'$ be the path obtained from rerouting $P$ through $Q$ and let $\mathcal{P}' = (\mathcal{P} - P) \cup \{P'\}$. It follows that $\mathcal{P}'$ is a path system equivalent to $\mathcal{P}$ and does not contain the edge $e$, contradicting our assumption of minimality.

It follows that $Q$ has an internal vertex contained in $\mathcal{P}$. Thus, there exists a dive $R'$ intersecting an internal vertex of $Q$. If $R'$ intersects $C_l$ as well, it follows that $R'$ has depth $l$ and that $\Delta_{R'}$ is strictly contained in $\Delta_R$, contradicting our choice of $R$. Otherwise, $R'$ has depth $l-1$, as desired by the claim. $\diamond$

Observe that for any two distinct dives $R$ and $R'$, any vertex $v \in V(R) \cap V(R')$ must be contained in $\Omega$.

We now finish the proof of the lemma. Lest the lemma hold, we may assume that there exists a dive of depth at least $s-t$. Consequently by Claim 3, for all $1 \le i \le s-t$, there exists a dive $R_i$ of depth $i$. By planarity, $R_i$ intersects $C_j$ for all $1 \le j \le i$, and the path $R_i - \Omega$ intersects $C_j$ for all $2 \le j \le i$. We conclude that the set $\{(R_i - \Omega) \cup C_i : 2 \le i \le 2w(k)+2\}$ is a bramble. To see the order of this bramble, first observe that the paths $R_i - \Omega$ and $R_j - \Omega$ are pairwise disjoint for all $2 \le i < j \le 2w(k)+2$. It now follows that every vertex $v$ is contained in at most two distinct subgraphs $(R_i - \Omega) \cup C_i$ for $2 \le i \le 2w(k)+2$. We conclude that there does not exist a cover of size $w(k)$, and consequently, the bramble has order at least $w(k)+1$. Theorem 5.13 and Claim 2 yield a contradiction, completing the proof of the lemma. $\square$

We now extend Lemma 5.14 to the "cylinder" case. The following lemma essentially shows that given an almost planar graph embedded in a cylinder with a large number of homotopic cycles and a path system with all it's endpoints contained in the boundary of the cylinder, then we can re-arrange the linkage so that only a bounded number of elements intersect a smaller middle portion of the cylinder. Moreover, given a cylindrical grid in this middle portion of the cylinder, we can ensure that the new path system follows the grid when traversing the middle portion of the grid.

**Lemma 5.15.** *Let $l$, $t$, and $k$ be positive integers. Assume Hypothesis 1. Let $w(k)$ be the value given by Corollary 5.12. Let $\mathcal{P}$ be a path system of order $k$, and assume that for every $P \in \mathcal{P}$ the endpoints of $P$ are contained in $V(H) \cup V(W_s)$. Let $\mathcal{R}$ be a linkage of order $t$ which is orthogonal to $\mathcal{C}$. If $s \ge 15w(k)^2 + l \ge 2(3(2w(k)+2)(w(k)) + (l+3w(k)) + 1$ and $t \ge 7w(k)$, then the following hold. There exists a linkage $\mathcal{R}' \subseteq \mathcal{R}$ of order at most $w(k)$, a path system $\mathcal{P}'$ equivalent to $\mathcal{P}$, and an index $i$ such that for every $P \in \mathcal{P}'$, the subgraph $P[V(W_i) \setminus V(W_{i+l})]$ is contained in $V(\mathcal{R}')$.*

*Proof.* Assume the lemma is false, and let $W$ along with the path systems $\mathcal{P}$ and $\mathcal{R}$ form a counterexample on a minimal number of edges. Subject to having a minimal number of edges, we pick an embedding up to 3-separations $(\{G_0, G_1, \ldots, G_m\}, \sigma, \{\Delta_1, \ldots, \Delta_m\})$ of $G$ which minimizes $m$. Observe that there does not exist an edge of $G_0$ which is not contained in $\mathcal{C} \cup \mathcal{P} \cup \mathcal{R}$ by our choice to minimize the number of edges. We will need to consider several different path systems throughout the proof, and in anticipation, fix $\mathcal{P}_1 = \mathcal{P}$.

We proceed in several steps. The first claim parallels Claim 1 in the proof of Lemma 5.14.

*Claim* 4. There does not exist $j \in \{1, \ldots, m\}$ with $G_j$ a subgraph of $G - V(W_s)$.

*Proof.* Note that no element of $\mathcal{P}_1$ has an endpoint contained in $V(G) \setminus (V(W_s) \cup V(C_1))$. Assume $j$ is an index such that $G_j$ is a subgraph of $G - V(W_s)$. At most one element $C_i$ intersects an edge of $G_j$, and similarly, at most one element of $\mathcal{R}$ intersects an edge of $G_j$. For all $R \in \mathcal{R}$ and $C \in \mathcal{C}$, we have that $R \cup C$ does not contain any cycle other than the cycle $C$. Thus, we see that there exists a spanning tree $T$ contained in $G_j$ such that both $\bigcup_1^s C_i \cap G_j$ and $\bigcup_{R \in \mathcal{R}} R \cap G_j$ are both subgraphs of $T$. If we consider how $\mathcal{P}_1$ can intersect the edges of $G_j$, we see that there exists at most one element $P$ of $\mathcal{P}_1$ which intersects an edge of $G_j$. If $P$ cannot be rerouted in $G_j$ to use only edges of $T$, we see that there exists an edge $e$ of $G_j$ and a path system $\mathcal{P}'_1$ equivalent to $\mathcal{P}_1$ such that for all $P' \in \mathcal{P}'_1$, $P' \cap G_j$ is a subgraph of $T + e$. Note here we are using the fact that no element of $\mathcal{P}_1$ has an endpoint in $V(G_j) \setminus V(G_0)$. For this reason, we are not able to prove the stronger statement that $m = 0$ and $G$ is planar, because some of the endpoints may be contained in $V(W_s)$.

If there exists an edge $f$ contained in $E(G_j) \setminus (E(T) \cup \{e\})$, we see that $J - f$ forms a counter-example on fewer edges. Thus we may assume that $G_j = T$ (or $T + e$ when the edge $e$ is defined). However, by embedding $T$ ($T + e$) in the disc $\Delta_j$ with the vertices of $V(G_0) \cap V(G_j)$ on the border, we see that $G_0 \cup G_j$ embeds in $\Delta - \left(\bigcup_{i \ne j} \Delta_i\right)$. As in Claim 1, we need to embed $T$ ($T + e$) with any vertices of $V(T) \cap V(\mathcal{C})$ on the boundary of the disc $\Delta_j$ in order to avoid altering the subgraphs $W_i$ for $1 \leq i \leq s$. Thus, we contradict our choice of embedding up to 3-separations to minimize $m$, proving the claim. $\diamond$

Let $W'$ be the subgraph $G - (V(W_{s-1}) - C_{s-1})$. Thus, $W'$ is the subgraph obtained from $G$ by deleting the portion of $G$ contained "inside" the disc bounded by $C_{s-1}$. By the previous claim, $W'$ is a subgraph of $G_0$ and embedded in the disc.

Let the elements of $\mathcal{R}$ be labeled $R_1, \ldots, R_t$ such that if we let $r_i$ be the endpoint of $R_i$ in $C_1$ for $1 \leq i \leq t$, we have the $r_1, r_2, \ldots, r_t$ occur on $C_1$ in that order. Define $W^*$ to be the subgraph $W_{3w(k)} - (V(W_{s-3w(k)}) - V(C_{s-3w(k)}))$. Let $R_i^* = R_{3w(k)+i}$ for $1 \leq i \leq w(k)$ and let $\mathcal{R}^*$ be the linkage $\{R_1^*, \ldots, R_{w(k)}^*\}$. Let $C^*$ be the unique cycle contained in $R_1^* \cup R_{w(k)}^* \cup ((C_{3w(k)} \cup C_{s-3w(k)}) - V(R_1))$ and let $(J^*, \Omega^*)$ be the embedding induced society of $C^*$. Finally, let $\mathcal{Q}$ be the linkage given by $\{C_i - V(J^*) : 3w(k) \leq i \leq s - 3w(k)\}$.

Pick a path system $\mathcal{P}_2$ equivalent to $\mathcal{P}_1$ satisfying the following.

i. $V(\mathcal{P}_2)$ is disjoint from $V(J^*)$.
ii. The graph $\bigcup_{Q \in \mathcal{Q}} Q \cup \bigcup_{P \in \mathcal{P}_2} P$ has as few vertices of degree at least 3 as possible.

To see that such a path system $\mathcal{P}_2$ exists, it suffices to show that there exists a path system satisfying *i*. Let $\mathcal{C}' = \{C_1', \ldots, C_{3w(k)}'\}$ be the planar nest with $C_i'$ the unique cycle contained in $R_{i+1} \cup R_{t-i} \cup (C_i - V(R_1)) \cup (C_{s-i} - V(R_1))$ for $1 \leq i \leq 3w(k)$. Note that since $C_i'$ is contained in $W'$, it is trivially grounded and by construction $\mathcal{C}'$ satisfies the definition of an $3w(k)$-nest. Moreover, if we let $(J', \Omega(J'))$ be the embedding induced society by $C_1'$, we see that the path system $\mathcal{P}_1$ has all it's endpoints disjoint from $J' - V(C_1')$. Note as well that $C_{3w(k)}' = C^*$. Thus by applying Lemma 5.14, we see that there exists a path system equivalent to $\mathcal{P}_1$ which is disjoint from the embedding induced society of $C_{3w(k)}' = J^*$, as desired. Note, we are assuming here that $w(k) \geq 3$ in order to simplify the constants.

*Claim* 5. The graph formed by $\bigcup_{Q \in \mathcal{Q}} Q \cup \bigcup_{P \in \mathcal{P}_2} P$ has treewidth strictly less than $w(k)$.

*Proof.* Let $A$ be the graph given by $\bigcup_{Q \in \mathcal{Q}} Q \cup \bigcup_{P \in \mathcal{P}_2} P$. Notice that by construction, every vertex of degree at least 3 in $A$ is a vertex of $\mathcal{P}_2$. If $tw(A) \geq w(k)$, then by Corollary 5.12, there exists a path system, call it $\mathcal{P}_2'$, equivalent to $\mathcal{P}_2$ contained in $A$ avoiding some vertex of degree at least 3. However, the graph $\bigcup_{Q \in \mathcal{Q}} Q \cup \bigcup_{P \in \mathcal{P}_2'} P$ will have strictly fewer vertices of degree at least three, contradicting our choice of $\mathcal{P}_2$ and proving the claim. $\diamond$

If every element $P$ of $\mathcal{P}_2$ could be divided into subpaths which each were perpendicular to the nest $\{C_{3w(k)}, \ldots, C_{s-3w(k)}\}$, then it would be an easy task to reroute each element through the subgraph $J^*$ so that it would follow $R_i^*$ for some $i$ when restricted to $W^*$ and prove the lemma. However, the paths $P \in \mathcal{P}_2$ are not necessarily so well behaved; the path $P$ may "bounce" around between the various cycles of $\{C_{3w(k)}, \ldots, C_{s-3w(k)}\}$. The next claim shows that these "bounces" are of bounded size.

We first need a definition to make explicit what we mean by "bounce". A *reversal* of $\mathcal{P}_2$ is a subpath $P$ of some element of $\mathcal{P}_2$ such that

i. $P$ is contained in $W^*$, and
ii. there exists an index $j$, $3w(k) \leq j \leq s - 3w(k)$ such that both endpoints of $P$ are contained in $V(C_j)$ and no internal vertex of $P$ is contained in $V(C_j)$.

The *depth* of a reversal $P$ with endpoints in $C_j$ is the maximum value of $|j - j'|$ such that $3w(k) \leq j' \leq s - 3w(k)$ and $V(P) \cap V(C_{j'}) \neq \emptyset$.

*Claim* 6. Every reversal of $\mathcal{P}_2$ has depth at most $2w(k)$.

*Proof.* The proof follows the proof of Claim 3. Let $P$ be a reversal of depth $n$ with endpoints contained in $C_j$ for some index $j$, and let $D_P$ be the subdisc of $\Delta$ bounded by $P$ and the subpath of $C_j - R_1^*$ containing the endpoints of $P$. We claim that there exists a reversal $P'$ contained in $D_P$ with endpoints in $C_j$ and depth $n - 1$. Assume not, and pick such a $P$ with $n$ minimal, and subject to that, with $D_P$ minimal by containment. By symmetry, we assume that $V(P) \cap V(C_{j'}) \neq \emptyset$ with $j' - j = n$. Inside $D_P$, there exists a subpath of $C_{j'-1}$ with both endpoints in $P$ and otherwise disjoint from $P$. Call it $Q$. If $Q$ were disjoint from $\mathcal{P}_2$, then we could reroute $\mathcal{P}_2$ through $Q$ to avoid some vertex of $C_{j'}$ and violate our choice of $\mathcal{P}_2$ to satisfy *ii*. We conclude that there exists a reversal $P'$ contained in $D_P$ with both endpoints in $C_j$ of depth at least $n-1$. If $P'$ has depth $n$, we violate our choice of to minimize $D_P$ by containment, and so we see that the desired reversal $P'$ of depth $n-1$ exists. Note that $P'$ is disjoint from $P$ except for possibly at its endpoints in $C_j$.

Now assume that there exists a reversal $R_{2w(k)+1}$ of depth $2w(k)+1$ with endpoints contained in $C_j$. By symmetry, we may assume that $R_{2w(k)+1}$ intersects the path $C_{j+2w(k)+1}$. We have just seen that there must exist reversals $R_i$ of depth $i$ for $1 \leq i \leq 2w(k)+1$, each with endpoints on $C_j$. Moreover, if we let $D_i$ be the disc bounded by $R_i$ and the subpath of $C_j - R_1^*$ connecting its endpoints, we have that $D_1 \subseteq D_2 \subseteq \cdots \subseteq D_{2w(k)+1}$. For all $1 \leq i < j \leq 2w(k)+1$, if $R_i$ intersects $R_j$, the intersection must lie on $C_j$. Thus, if we let $R_i' = R_i - V(C_j)$ for $1 \leq i \leq 2w(k)+1$, we have that $V(R_i') \cap V(R_j') = \emptyset$ for $i \neq j$. By planarity, it follows that the elements of the set $\{R_i' \cup C_{j+i} : 1 \leq i \leq 2w(k)+1\}$ are pairwise intersecting and form a bramble. Every vertex is in at most two distinct elements of the form $R_i' \cup C_{j+i}$, and consequently, the bramble has order $w(k)+1$ in $\mathcal{P}_2 \cup \mathcal{Q}$, a contradiction to Theorem 5.13 and Claim 5. $\diamond$

We will now see that the elements of $\mathcal{P}_2$ can be subdivided into components which are nearly perpendicular to a subset of the cycles $C_{3w(k)}, \ldots, C_{s-3w(k)}$. First, we give the following definition. Let $\mathcal{S}$ be a path system, and let $S \in \mathcal{S}$ be an element with at least one internal vertex. Let the ends of $S$ be $x$ and $y$, and let $v$ be an internal vertex. We say that the path system $\mathcal{S}'$ is obtained by *subdividing the element S of $\mathcal{S}$ at the vertex v* if $\mathcal{S}' = (\mathcal{S} - S) \cup \{xSv, vSy\}$. The path system $\mathcal{S}'$ is a *refinement of $\mathcal{S}$* if $\mathcal{S}'$ is obtained by repeatedly subdividing elements of $\mathcal{S}$.

We fix the set of cycles for $\mathcal{C}' = \{C_{i(3w(k))} : 2 \leq i \leq 2w(k)+2\} \cup \{C_{s-i3w(k)} : 2 \leq i \leq 2w(k)+2\}$. We note that by our assumption on $s$ that $s \geq 12(w(k))^2$. Thus cycles in $\{C_{i(3w(k))} : 2 \leq i \leq 2w(k)+2\}$ and cycles in $\{C_{s-i3w(k)} : 2 \leq i \leq 2w(k)+2\}$ are disjoint.

*Claim* 7. There exists a refinement $\mathcal{P}_3$ of $\mathcal{P}_2$ which is nearly perpendicular to the set of cycles $\mathcal{C}'$. The order of $\mathcal{P}_3$ is at most $w(k)+k$, and at most $w(k)$ elements of $\mathcal{P}_3$ intersect $C_{6w(k)}$.

*Proof.* We pick a set of vertices $X \subseteq V(\mathcal{P}_2)$ satisfying the following properties.

  i. For all $x \in X$, $x \in V(C_{6w(k)})$.
  ii. For all $P \in \mathcal{P}_2$ and $x, y \in V(P) \cap X$, there exists a vertex $z$ on the subpath $xPy$ such that $z \in V(C_{3w(k)}) \cup V(C_{s-3w(k)})$.
  iii. Subject to *i* and *ii*, the set $X$ is chosen with $|X|$ maximal.

We now define a set $Z$ as follows. For every $P \in \mathcal{P}_2$ such that there exists distinct $x, y \in X \cap V(P)$, there exists a vertex $z \in Z$ such that $z \in V(xPy) \cap (V(C_{3w(k)}) \cup V(C_{s-3w(k)}))$. Moreover, we pick $Z$ to be minimal over all such sets. Thus, the set $X$ can be thought of as selecting a vertex $x$ for each time the path system $\mathcal{P}_2$ returns to the cycle $C_{6w(k)}$ after first visiting one of the "outside" cycles $C_{3w(k)}$ or $C_{s-3w(k)}$. The set $Z$ then consists of vertices on the cycles $C_{3w(k)}$ and $C_{s-3w(k)}$ separating any pair of vertices of $X$ contained in the same element of $\mathcal{P}_2$.

Let $\mathcal{P}_3$ be the refinement of $\mathcal{P}_2$ obtained by subdividing the elements of $\mathcal{P}_2$ at the vertices of $Z$. We claim that $\mathcal{P}_3$ is the refinement desired by the claim. First, we see that $\mathcal{P}_3$ is nearly perpendicular to the set of cycles $\mathcal{C}'$. Property *ii* in the definition follows immediately, as if we had vertices $x, y, z$ violating *ii*, it would yield a reversal of depth $3w(k)$, contradicting Claim 6. To see that Property *i* holds, first observe that for all $P \in \mathcal{P}_3$, $P$ contains a vertex in $C_{6w(k)}$. Moreover, $P$ has no endpoint in $W_{3w(k)} - W_{s-3w(k)}$. It follows by the planarity $W_{3w(k)} - W_{s-3w(k)}$ and the fact that $P$ cannot contain a reversal of order $3w(k)$ that $P$ intersects every element $\mathcal{C}'$, proving that $\mathcal{P}_3$ satisfies *ii*.

We now show that the order of $\mathcal{P}_3$ satisfies the desired bounds. Assume that there exist at least $w(k)+1$ elements of $\mathcal{P}_3$ which intersect $C_{6w(k)}$. If we let $S$ be the set of endpoints of $\mathcal{P}_3$, we see that $\{P - S : P \in \mathcal{P}_3\}$ contains a linkage $\mathcal{P}_3'$ of order $w(k)+1$ such that $\mathcal{P}_3'$ is nearly perpendicular to $\{C_{3i(w(k))} : 2 \leq i \leq 2w(k)+2\}$. However, the sets $\mathcal{B} = \{P \cup (C_{i3w(k)} - V(J^*)) : P \in \mathcal{P}_3', 1 \leq i \leq 2w(k)+2\}$ forms a bramble of order $w(k)+1$ which is contained in $\mathcal{P}_3 \cup \mathcal{Q}$, contradicting Claim 5. Thus, we see that at most $w(k)$ elements of $\mathcal{P}_3$ intersect $C_{6w(k)}$. The bound on the order of $\mathcal{P}_3$ now follows from the fact that any element of $\mathcal{P}_3$ which does not intersect $C_{6w(k)}$ is an element of $\mathcal{P}_2$ as well, and there are at most $k$ such elements. ◇

Fix $\mathcal{P}_3$ as in Claim 7. We now pick an appropriate path system $\mathcal{P}_4$ which is equivalent to $\mathcal{P}_3$. First, we give some notation. Let $t'$ be a non-negative integer. Let $W_{t'}^*$ be the subgraph $W_{3(2t'+2)w(k)} - (W_{s-3(2t'+2)w(k)} - C_{s-3(2t'+2)w(k)})$, and let $J_{t'}^*$ be the subgraph $J^* \cap W_{t'}^*$. We let

$$\mathcal{C}_{t'}' = \{C_{3iw(k)} : 2t'+2 \leq i \leq 2w(k)+2\} \cup \{C_{s-3iw(k)} : 2t'+2 \leq i \leq 2w(k)+2\}.$$

Thus $\mathcal{C}_0' = \mathcal{C}'$. We now fix $\mathcal{P}_4$ to be a path system equivalent to $\mathcal{P}_3$ and fix non-negative integer $t'$ satisfying the following properties.

i. $J_{t'}^* \cap \left(\bigcup_{P \in \mathcal{P}_4} P\right)$ is equal to $J_{t'}^* \cap \left(\bigcup_1^{t'} R_i^*\right)$.
ii. There exist at most $w(k) - t'$ elements which intersect $W_{t'}^* - J_{t'}^*$ and every such an element is nearly perpendicular to $\mathcal{C}_{t'}'$.
iii. No element of $\mathcal{P}_4$ intersects both $J_{t'}^*$ and $W_{t'}^* - V(J_{t'}^*)$.

Moreover, we pick $\mathcal{P}_4$ and $t'$ over all such possibilities to maximize the value of $t'$. Note that such a path system $\mathcal{P}_4$ exists, as $\mathcal{P}_3$ satisfies *i-iii* with $t' = 0$ since by Claim 7 at most $w(k)$ elements of $\mathcal{P}_3$ intersect the elements of $\mathcal{C}'$. The next claim will essentially complete the proof of the lemma.

*Claim* 8. There do not exist an element of $\mathcal{P}_4$ which intersects the subgraph $W_{t'}^* - V(J_{t'}^*)$.

*Proof.* Assume the claim is false. We will derive a contradiction to our choice of $\mathcal{P}_4$ to maximize $t'$. Observe that $t' < w(k)$. Fix a vertex $v$ in $R_{t'+1}^* \cap C_{s-3(2t'+2)w(k)}$, and let $P \in \mathcal{P}_4$ be the first path we encounter when traversing $C_{s-3(2t'+2)w(k)}$ starting from $v$ and moving away from the vertices of $\bigcup_1^{t'} (R_i^* \cap C_{s-3(2t'+2)w(k)})$. Let $u$ be a vertex of $P \cap C_{s-3(2t'+2)w(k)}$, and let $S$ be the subpath of $C_{s-3(2t'+2)w(k)}$ with endpoints $u$ and $v$ intersecting $R_{w(k)}^* \cap C_{s-3(2t'+2)w(k)}$. Let $v'$ be a vertex of $R_{t'+1} \cap C_{3(2t'+2)w(k)}$. Let $u'$ be a vertex of $P \cap C_{3(2t'+2)w(k)}$ and let $S'$ be the subpath of $C_{3(2t'+2)w(k)}$ linking $u'$ and $v'$ intersecting $R_{w(k)}^* \cap C_{s-3(2t'+2)w(k)}$. Finally, let $D^*$ be the subgraph of $W^*$ contained in the disc bounded by the paths $R_{t'+1}^* \cap J_{t'}^*$, $P$, $S$ and $S'$.

We claim that there does not exist $P' \in \mathcal{P}_4$, $P' \neq P$, and index $j$, $3(2t'+3)w(k) \leq j \leq s - 3(2t'+2)w(k)$ such that $C_j \in \mathcal{C}_{t'}'$ and $P'$ intersects $C_j \cap D^*$. Assume otherwise. As $P'$ does not intersect any of the paths $R_{w(k)}^* \cap J_{t'}^*$, $P \cap D^*$, or $S$ by construction, and by the planarity of $D^*$, we see that there exist vertices $x$ and $y$ on $P'$ such that $x, y \in V(P') \cap V(C_{j-3w(k)})$ and the subpath $xP'y$ intersects $C_j$. However, this contradicts the fact that $\mathcal{P}_4$ is nearly perpendicular to $\mathcal{C}_{t'}'$. We conclude that no such $P'$ exists.

Let $S''$ be a subpath of $C_{3(2t'+3)w(k)}$ intersecting $R_{w(k)}^*$ and linking a vertex of $R_{t'+1}^* \cap C_{3(2t'+3)w(k)}$ and a vertex of $P \cap C_{3(2t'+3)w(k)}$. We have just seen that $S''$ must be disjoint from all elements of $\mathcal{P}_4$ except for $P$. Let $P''$ be the path obtained by rerouting $P$ through the path $S \cup S'' \cup (R_{t'+1}^* \cap J_{t'}^*)$. It now follows that if we let $\mathcal{P}_4'$ be the path system $(\mathcal{P}_4 - P) \cup P''$, we satisfy *i - iii* above with the integer $t' + 1$, contradicting our choice of $\mathcal{P}_4$ and $t'$. This contradiction completes the proof of the claim. ◇

The path system $\mathcal{P}_4$ is equivalent to a refinement of $\mathcal{P}_2$, which itself is equivalent to the original path system $\mathcal{P}_1$. Consequently, $\mathcal{P}_4$ contains a path system $\mathcal{P}_5$ which is equivalent to $\mathcal{P}_1$. The path system $\mathcal{P}_5$ satisfies $\bigcup_{P \in \mathcal{P}_5} P \cap W_{t'}^* = \bigcup_1^{t'} R_i^* \cap W_{t'}^*$. Moreover, by our choice of the original $s$, we see that $s - 2(3(2t'+2)w(k)) \geq l$, and so $W_{t'}^*$ contains $l$ consecutive cycles $C_i, C_{i+1}, \ldots, C_{i+l}$, as desired. □

### 5.6 Proof of Theorem 5.9

In this section, we give the proof Theorem 5.9, completing the proof of Theorem 3.3.

Intuitively, we fix a copy of a topological minor which has as few endpoints in the flat *r*-wall decomposition as possible. Let the topological minor be given by a path system $\mathcal{P}$. We can always find a large belt of the *r*-wall that has the cylindrical grid structure and does not contain any of the endpoints of $\mathcal{P}$. If indeed none of the endpoints are contained in the inside ring of the cylindrical grid as well, then we apply Lemma 5.14 and find a copy of the topological minor which doesn't contain the middle vertices of the $r$-wall decomposition, the desired outcome. Thus, we reduce to the case when some of the endpoints of $\mathcal{P}$ are contained in the inside ring. We then apply Lemma 5.15 and use property $v$. to move at least one of the endpoints outside the cylinder, contradicting our choice of $\mathcal{P}$ and completing the proof.

*Proof (Theorem 5.9).* Let $\delta$ and $t$ be given. Let $(A, G, H, X, W)$ be the given nearly flat $r$-wall decomposition with apex bound $t$. Let $w(\delta + t)$ be the value of the function in Theorem 5.12. We let

$$m = 2(2(\delta + t))^2 \left(15[w(\delta + t)]^2 + 2l + w(\delta + t)\right).$$

We fix a brick containing the vertex $v_r^{r/2}$, and let $\{C_1, C_2, \ldots, C_m\}$ be the $m$-target centered at this brick. We assume $r \geq 2m$, ensuring that the cycles are defined.

Fix a rooted graph $S$ in the $\delta$-folio. For any given model of $S$ in $J$, we may assume that it is given as a path system $\mathcal{P}_S$. We let $\bar{\mathcal{P}}$ be the refinement of $\mathcal{P}_S$ obtained by including as a terminal any vertex of $A$ which is an internal vertex of some element of $\mathcal{P}_S$. We let $\mathcal{P}$ be the path system given by $\{P - A : P \in \bar{\mathcal{P}}\}$. Note that $\mathcal{P}$ has at most $\delta + t$ elements. We define the values $n_i$ for $1 \leq i \leq 2(\delta + t)$ as follows,

$$n_i = i\left(15[w(\delta + t)]^2 + 2l + w(\delta + t)\right).$$

We are now able to give our requirements for $l$ and $d$.

$$l = \delta + t \text{ and } d = 2(\delta + t)n_{2(\delta + t)}.$$

Let $G(i)$ be the embedding induced society of $C_i$ for $1 \leq i \leq m$. For an index $j$, $0 \leq j \leq 2(\delta + t)$, we define $a_j = a_j(\mathcal{P})$ to be the maximum index $i$, $1 \leq i \leq m$, such that

i. There are exactly $j$ distinct endpoints of elements of $\mathcal{P}$ which are contained in $G(a_j)$,

ii. $a_j \leq m - 14w(\delta + t)l$, and

iii. subject to i and ii, $G(a_j - n_j) - G(a_j)$ contains no endpoint of any element of $\mathcal{P}$.

It is not necessarily the case that $a_j$ will be defined for every value of $j$. However, as we will see below, for any path system $\mathcal{P}$ arising from a given model of $S$ in $J$, there exists at least one index $j$ such that $a_j$ is defined and the value is bounded by a function of $\delta$ and $t$.

*Claim* 1. There exists an integer $j$ such that the value $a_j$ is defined. Moreover, for all $j$ for which $a_j$ is defined, then there exists an index $j' \leq j$ such that $a_{j'}$ is defined and $a_{j'} \leq m - 2(\delta + t)n_{2(\delta + t)}$.

*Proof.* The subgraphs $G((i-1)n_{2(\delta + t)}) - G(in_{2(\delta + t)})$ for $1 \leq i \leq 2(\delta + t) + 1$ are disjoint. As the linkage $\mathcal{P}$ has at most $\delta + t$ elements, it has at most $2(\delta + t)$ distinct endpoints. Thus, there exists an index $i'$ and value $j$ such that $G((i' - 1)n_{2(\delta + t)}) - G(i'n_{2(\delta + t)})$ is disjoint from the set of endpoints of elements of $\mathcal{P}$ and $G(i'n_{2(\delta + t)})$ contains exactly $j$ endpoints of elements of $\mathcal{P}_S$. Note that $(2(\delta + t) + 1)n_{2(\delta + t)} \leq m - 15w(\delta + t)^2 \leq m - 14w(\delta + t)$, and so ii in the definition of $a_j$ is satisfied. We conclude that $a_j$ is defined.

The same argument shows that if $a_j$ is defined for some index $j$, then there exists a value $j' \leq j$ such that $a'_j$ is defined and $a_{j'} \leq m - 2(\delta + t)n_{2(\delta + t)}$. ◇

We now fix the path system $\mathcal{P}_S$ forming a model of $S$ such that $a_j$ is defined for the path system $\mathcal{P}$ and the value $j$ is minimal over all such path systems and choices of $a_j$. By the previous claim, we see that $a_j \geq m - 2(\delta + t)n_{2(\delta + t)}$. To help keep the notation simple, let $a = a_j$. Let $Z$ be the set of endpoints of elements of $\mathcal{P}$. The next claim is the crux of our proof of Theorem 5.9. In the claim, we pick a path system forming a model of $S$ moving at least one of the vertices of $Z \cap G(a)$ further outside the $m$-nest, and thus derive a contradiction to $j$ minimal.

*Claim* 2. The value $j$ is 0, i.e. $Z$ is disjoint from $G(a)$.

*Proof.* Our first goal will be to find a path system $\mathcal{Q}_1$ contained in $J - A$ which is equivalent to $\mathcal{P}$ and intersects nicely with the nest $\mathcal{C} = \{C_{a-1}, \ldots, C_{a-n_j}\}$. Let $\mathcal{R}$ be a linkage of order $7w(\delta + t)$ which is orthogonal to $\mathcal{C}$ comprised of subpaths of the horizontal paths of the wall $W$. We additionally require that if $R, R' \in \mathcal{R}$ are subpaths of the $i$-th and $i'$-th horizontal paths of $W$, then $|i - i'| \geq 2l$; that is, we choose horizontal paths which are pairwise separated by at least $2l$ other horizontal paths of $W$ not included in $\mathcal{R}$. Note that to ensure the existence of such a linkage $\mathcal{R}$, we use the upper bound on $a$ given by *ii*.

Apply Lemma 5.15 to the nest $\mathcal{C}$ and the orthogonal linkage $\mathcal{R}$ to get a path system $\mathcal{Q}_1$ equivalent to $\mathcal{P}$ satisfying the statement of the Lemma 5.15. Fix the index $a'$ such that every subpath of $\mathcal{Q}_1$ contained in $G(a' - (n_{j-1} + 2l + w(\delta + t))) - G(a')$ is contained in the linkage $\mathcal{R}$. Note that we may assume that $a' \geq a - 15[w(\delta + t)]^2$. After first finding a path system $\mathcal{Q}_2$ equivalent to $\mathcal{Q}_1$ which eliminates a technicality, the remainder of the proof will proceed as follows. We find a pattern contained in $\mathcal{Q}_2$ centered at $C_{a'}$ of depth $l$ which can be replaced by a pattern contained in $G(a' - (n_{j-1} + 2l)) - G(a' - n_{j-1})$, and thus move at least one endpoint of the path system "outside" the cycle $G(a' - n_{j-1})$. Thus, we will contradict our choice of $\mathcal{P}$.

First, we consider the technicality mentioned above. We eliminate the possibility that elements intersect the graph $G(a')$ "needlessly". Consider an element of $\mathcal{Q}_1$, and let $Q$ be a component of the restriction to the vertex set of $G(a')$ such that $Q$ has both endpoints contained in $V(C_{a'})$. If we consider the embedding of $G(a')$ in the disc, then the path $Q$ divides the disc into two sub-discs $\Delta_1(Q)$ and $\Delta_2(Q)$. We say that $Q$ is *wasteful* if at least one of the discs $\Delta_i(Q)$ does not contain a vertex of $Z$. We claim that there exists a path system $\mathcal{Q}_2$ equivalent to $\mathcal{Q}_1$ such that $\mathcal{Q}_2$ intersects $G(a' - (n_{j-1} + 2l)) - G(a')$ exactly in the a subset of at most $w(\delta + t)$ components of $\mathcal{R}$, and, moreover, $\mathcal{Q}_2$ does not contain a wasteful path $Q$. To see this, consider a wasteful path $Q$ in $\mathcal{Q}_1$ and assume that $\Delta_1(Q)$ does not contain a vertex of $Z$. Assume we pick such a $Q$ to minimize $\Delta_1(Q)$ by containment. It follows that no other element of $\mathcal{Q}_1$ intersects $\Delta_1(Q)$. To see this, such a component cannot contain an element of $Z$ by the choice of $Q$. Furthermore, if there exists a component $Q'$ intersecting $\Delta_1(Q)$, then $Q'$ must be wasteful and we violate the choice of $Q$ to minimize $\Delta_1(Q)$. We conclude, by rerouting $Q$ through a subpath in the cycle $C_{a'-(n_{j-1}+2l+w(\delta+t))}$, we can find a path system equivalent to $\mathcal{Q}_1$ which satisfies the property that the intersection with the subgraph $G(a' - (n_{j-1} + 2l + w(\delta + t) - 1)) - G(a')$ is contained in $\mathcal{R}$ and which has one less wasteful subpath than $\mathcal{Q}_1$. Thus, by inductively iterating this process at most $w(\delta + t)$ times, we arrive at the desired path system $\mathcal{Q}_2$.

Consider a component $X$ of the graph $\bigcup_{Q \in \mathcal{Q}_2} Q \cap G(a' - l)$ such that $V(X) \cap Z \neq \emptyset$, and let $T$ be the path system associated to the graph $X$. There possibly exist edges of the original path system $\bar{\mathcal{P}}$ with one endpoint in $V(T) \cap Z$ and the other endpoint in $A$. Let $T^+$ be the union of $T$ and all such edges. By construction $T^+$ is a pattern centered at $C_{a'}$ of depth $l$.

The patterns $T^+$ come in two slightly different types: either $T^+$ can intersect $C_{a'-l}$, or alternatively, the path system $T$ could be entirely contained in $G(a')$. If such a $T$ exists of the second type, we fix $T$ to be such a path system, and fix $\Delta$ to be a subdisc of the embedded graph $G(a' - l)$ containing $T$ and otherwise not intersecting $V(\mathcal{Q}_2) \setminus V(T)$.

Alternatively, we consider the case when every choice of $T$ must intersect $V(C_{a'-l})$. We claim that $T^+$ can be chosen so that there exists a subpath $L$ of $C_{a'-l}$ such that $L$ contains all the vertices of $V(T^+) \cap V(C_{a'-l})$ and $L$ is otherwise disjoint from $V(\mathcal{Q}_2)$. To ensure that $L$ is unique for every $T^+$, we fix an edge $e \in E(C_{a'-l}) \setminus E(\mathcal{R})$, and pick the path $L$ so that it does not contain the edge $e$. For every such $T^+$, fix $L(T)$ to be a minimal subpath of $C_{a'-l}$ containing all the vertices of $V(T^+) \cap V(C_{a'-l})$ and not containing the edge $e$. Fix $T$ such that $L(T)$ is minimal by containment. Given the embedding up to 3-separations of $G(a' - l)$ in the disc, there exists a subdisc $\Delta$ containing the vertices of $T$ such that every vertex in the boundary of $\Delta$ is either contained in $T$ or in $L(T)$. If there existed a $T'$ intersecting the disc $\Delta$, then every vertex of $V(T') \cap V(C_{a'-l})$ would be contained in $L(T)$ by planarity. Thus, $L(T')$ would be a proper subpath of $L(T)$, a contradiction. Alternatively, if some subpath $Q$ of $\mathcal{Q}_2$ not contained in $T$ intersects $\Delta$, then as we have just seen, $Q$ cannot contain any vertex in $Z$. Therefore, the existence of $Q$ implies the existence of a wasteful path, again a contradiction. We conclude that $\Delta$ intersects $\mathcal{Q}_2$ only in the vertices of $V(T)$.

We now will replace the pattern $T^+$ by a homeomorphic pattern contained in $G(a' - (n_{j-1} + 2l)) - G(a' - n_{j-1})$ to find a new path system $\mathcal{Q}^*_S$ forming a model of $S$ in $J$. Moreover, if we construct $\bar{\mathcal{Q}}^*$ and $\mathcal{Q}^*$ analogously to $\bar{\mathcal{P}}$ and $\mathcal{P}$, we see that $a_{j-1}(\mathcal{Q}^*)$ will be defined, contradicting our choice of $\mathcal{P}$ to minimize $j$.

We first consider the case when $T$ does not contain any vertex of $C_{a'-l}$. In this case, we pick a brick $B$ in $G(a'-(n_{j-1}+2l))-G(a'-n_{j-1})$ which does not intersect $\mathcal{R}$. By property $v$, the brick $B$ admits an $l$-target and so contains a pattern $\bar{T}^+$ homeomorphic to $T^+$. In fact, as the pattern $T^+$ does not intersect $C_{a'-l}$, we see that $V(\bar{T}^+) \setminus A$ is contained in the embedding induced society of the brick $B$. Thus, $V(\bar{T}^+)$ is disjoint from the linkage $\mathcal{R}$, and consequently, from $V(\mathcal{Q}_2)$. Let $\mathcal{Q}_S^*$ be a model for $S$ in $J$ obtained from $\mathcal{Q}_2$ by deleting the vertices of $V(T)$ and adding the corresponding paths in $\bar{T}^+$.

We now consider the alternative case when at least one element of $T^+$ intersects $C_{a'-l}$. We let $\bar{\mathcal{R}}$ be the subset of elements $R$ of $\mathcal{R}$ such that $R$ is contained in some element of $\mathcal{Q}_2$ and $R$ has an endpoint in $L(T^+)$. As we have seen in the above paragraph, for every $R \in \bar{\mathcal{R}}$, $R$ has one endpoint in $V(T)$. Let $R_1$ and $R_2$ be elements of $\bar{\mathcal{R}}$ containing the endpoint of $L(T^+)$. We now turn our attention to the graph $G(a'-(n_{j-1}+2l))$ and it's embedding up to 3-separations in the disc. By construction, there exists a subdisc of the embedding, call it $\bar{\Delta}$ such that $\bar{\Delta}$ is bounded only by vertices $\bar{\mathcal{R}}$, $T$, and a subpath of the cycle $C_{a'-(n_{j-1}+2l)}$. We also assume that the disc $\bar{\Delta}$ in fact contains all the elements of $\bar{\mathcal{R}}$ containing the endpoint of $L(T^+)$. If we look at the subgraph of $G(a'-(n_{j-1}+2l))-G(a'-n_{j-1})$ contained in $\bar{\Delta}$, then by property $v$ of Theorem 5.8 this subgraph contains an $l$-target centered at a brick and a pattern $\bar{T}^+$ homeomorphic to $T^+$. We may assume, in fact, that for any element $P \in T^+$ such that $P$ has an endpoint contained in $R \in \bar{\mathcal{R}}$, the corresponding element $\bar{P}$ in $\bar{T}^+$ has as the other endpoint a vertex of $R \cap V(C_{(a'+n_{j-1}+2l)})$. Thus, by deleting all the vertices of $\mathcal{Q}_2$ contained in $\bar{\Delta}-V(C_{(a'+n_{j-1}+2l)})$ from $\mathcal{P}_S$ and adding $\bar{T}^+$, we find a path system $\mathcal{Q}_S^*$ forming a model of $S$ in $J$.

Given the path system $\mathcal{Q}_S^*$, we define $\bar{\mathcal{Q}}^*$ and $\mathcal{Q}^*$ as in the definition of $\bar{\mathcal{P}}$ and $\mathcal{P}$. In either of the two cases above, we conclude that the constructed path system $\mathcal{Q}^*$ has at most $j-1$ endpoints contained in the subgraph $G(a')$ and no endpoint contained in $G(a'-n_{j-1})-G(a')$, contradicting our choice of $\mathcal{P}$ and proving the claim. Note that we are using here the property that there are no roots of $J$ contained in $G$ to ensure that the new path system $\mathcal{Q}_S^*$ does in fact form a model of the (rooted) topological minor of $S$. $\diamond$

Theorem 5.9 now follows by the upper bound on $a$ and Lemma 5.14. As the graph $G(a)$ contains the nest $\{C_m, \ldots, C_{m-a}\}$ and no endpoint of the path system $\mathcal{P}$, there exists a path system $\mathcal{P}^*$ in the subgraph $G \cup H$ which is equivalent to $\mathcal{P}$ and does not contain any vertex of $G(m)$. Specifically, by extending $\mathcal{P}^*$ to a model of $S$ in $J$ using edges incident the apex set $A$, we see that there is a model of $S$ which does not contain the peg $v_r^{r/2} \in V(G(m))$, as claimed. $\square$

## 6 Immersion

Let $G, H$ be graphs. An *immersion* of $H$ in $G$ is a function $\alpha$ with domain $V(H) \cup E(H)$, such that:

- $\alpha(v) \in V(G)$ for all $v \in V(H)$, and $\alpha(u) \neq \alpha(v)$ for all distinct $u, v \in V(H)$,
- for each edge $e$ of $H$, if $e$ has distinct ends $u, v$ then $\alpha(e)$ is a path of $G$ with ends $\alpha(u), \alpha(v)$, and if $e$ is a loop incident with a vertex $v$ then $\alpha(e)$ is a cycle of $G$ with $\alpha(v) \in V(\alpha(e))$, and
- for all distinct $e, f \in E(H), E(\alpha(e) \cap \alpha(f)) = \emptyset$.

In fact, we may impose on another condition in the definition of immersion, that

- for all $v \in V(H)$ and $e \in E(H)$, if $e$ is not incident with $v$ in $H$ then $\alpha(v) \notin V(\alpha(e))$.

Let us call this "strong immersion".

In this section, we show that our main theorem, Theorem 1.1 implies that the immersion containment problem is also fixed-parameter tractable parameterized by the order of $|E(H)|$. However, our reduction from Theorem 1.1 does not work for the "strong immersion" containment problem. We conjecture that the strong immersion containment is fixed-parameter tractable parameterized by the order of $|E(H)|$, too.

**Theorem 6.1.** *For every fixed graph $H$, there is a $O(|V(G)|^3)$ time algorithm that decides if $H$ is an immersion in $G$.*

*Proof.* Let $k = |E(H)| + |V(H)|$. We construct first a new graph $G'$ from $G$ by subdividing each vertex and replacing each original vertex by $k$ duplicates. Formally, for each $e \in E(G)$, there is a vertex $e'$ in $G'$; for each vertex $v \in V(G)$, there are $k$ vertices $v_1, \ldots, v_k$ in $G'$, and if $v \in V(G)$ is an endpoint of $e \in E(G)$, then vertex $e' \in V(G')$ is adjacent to $v_1, \ldots, v_k$ in $G'$. Note that the degree of $e'$ is $2k$.

Let $\ell = 2k|V(H)| + 1$ and let us use the algorithm of Theorem 1.1 to find a $K_\ell$ topological minor in $G'$. We claim that if there is such a topological minor model $\phi : V(K_\ell) \to V(G')$, then $H$ has an immersion in $G$. To see this, observe first that $\phi(v)$ is a vertex with degree at least $\ell - 1 > 2k$, thus $\phi(v) = u_i$ for some $u \in V(G)$; let us define $\alpha(v) = u$ in this case. It is clear that $\alpha$ maps at most $k$ vertices of $H$ to the same vertex of $G$. As $\ell/k > |V(H)|$ holds, one can select vertices $x_1, \ldots, x_{|V(H)|}$ whose images in $\phi$ are all distinct. For any $1 \le i, j \le |V(H)|$, the path $\phi(x_i x_j)$ between $\phi(x_i)$ and $\phi(x_j)$ in $G'$ gives a path $\alpha(x_i x_j)$ between $\alpha(x_i)$ and $\alpha(x_j)$ in a natural way. As the paths $\phi(x_i x_j)$ are pairwise internally vertex disjoint in $G'$, the paths $\alpha(x_i x_j)$ are pairwise edge disjoint in $G$: a vertex $e' \in E(G')$ can be used by at most one of the paths $\phi(x_i x_j)$. Therefore, $\phi$ shows that $K_{|V(H)|}$ has an immersion in $G$, which immediately implies that $H$ has an immersion in $G$. This means that we are done in the case when $K_\ell$ is a topological minor of $G'$.

Suppose now that $K_\ell$ is not a topological minor of $G'$. We modify $G'$ to obtain a new graph $G''$ as follows. For every $v \in V(G)$, we introduce a new copy of $K_\ell$ and identify $v_1$ with a vertex of $K_\ell$. Thus the number of vertices of $G''$ is $|V(G')| + |V(G)|(\ell - 1)$. Similarly, we obtain $H''$ from $H$ by introducing for each $u \in V(H)$ a new copy of $K_\ell$ and identifying $u$ and a vertex of $K_\ell$ (so $|V(H'')| = \ell|V(H)|$).

We claim that $H''$ is a topological minor of $G''$ if and only if $H$ has an immersion in $G$. For the if part, suppose that $\alpha$ is an immersion of $H$ in $G$. In this case, it is easy to construct a model $\phi$ of $H''$ in $G''$: if $\alpha(u) = v$ for some $u \in V(H)$ and $v \in V(G)$, then we set $\phi(u) = v_1$, map the clique attached to $v$ in $H''$ to the clique attached to $v_1$, and transform each path $\alpha(u_1 u_2)$ in $G$ into a corresponding path $\phi(u_1 u_2)$ in $G''$. We can ensure that the paths in $\phi$ are internally vertex disjoint: the paths in $\alpha$ are edge disjoint (so we can ensure that each vertex $e' \in V(G'')$ is used at most once) and the $k$ vertices $v_1, \ldots, v_k$ in $G''$ are sufficient to accommodate the at most $|E(H)|$ paths going through $v$ in $\alpha$.

For the only if part, suppose that $\phi$ is a model of $H''$ in $G''$. Consider a vertex $u$ of $H''$ that also appears in $H$ (i.e., it is not a vertex introduced by a new clique). The degree of $u$ in $H''$ is more than $\ell - 1$ (assuming that $H$ has no isolated vertices) and $u$ is part of an $\ell$-clique in $H''$. Thus $\phi(u)$ is a vertex of $G''$ having degree more than $\ell - 1$ and part of a topological minor model of a $\ell$-clique. We claim that $\phi(u) = v_1$ for some $v \in V(G)$. Every model of an $\ell$-clique is fully contained in a biconnected component of $G''$. As $G'$ has no $\ell$-clique topological minor, such a biconnected component must be one of the $K_\ell$-cliques created in the construction of $G''$. Furthermore, the new vertices of such a clique have degree exactly $\ell - 1$, thus $\phi(u)$ can be only a vertex $v_1$ for some $v \in V(G)$. Thus $\phi$ restricted to $H$ is a topological minor model of $H$ that does not go inside the cliques, which means that it is a topological minor model of $H$ in $G'$. Arguing as in the first part of the proof, it follows that $H$ has an immersion in $G$.

Let us estimate the running time of the algorithm. First, we can assume that $|E(G)| \le c_H |V(G)|$ for some constant $c_H$ depending only on $H$: by a classical result of Mader, if the average degree of $G$ is sufficiently large, then $G$ has a $K_{|V(H)|}$ topological minor, immediately implying that $H$ has an immersion in $G$. Therefore, the number of vertices of $G'$ is $k|V(G)| + |E(G)| = O(|V(G)|)$ (for fixed $H$). The construction of $G''$ increases the number of vertices by a factor of $\ell$, hence $|V(G'')| = O(|V(G)|)$ also holds. Thus both invocation of Theorem 1.1 needs $O(|V(G)|^3)$ time. □